\documentclass[fleqn,10pt]{wlscirep}
\usepackage[utf8]{inputenc}
\usepackage[T1]{fontenc}
\usepackage{hyperref} 
\title{Single-exposure x-ray dark-field imaging: quantifying sample microstructure using a single-grid setup}
\author[1,*]{Ying Ying How}
\author[1]{David M. Paganin}
\author[1]{Kaye S. Morgan}
\affil[1]{School of Physics and Astronomy, Monash University, Clayton, VIC 3800, Australia}

\affil[*]{Ying.How1@monash.edu}

\begin{abstract}
The size of the smallest detectable sample feature in an x-ray imaging system is usually restricted by the spatial resolution of the system. This limitation can now be overcome using the diffusive dark-field signal, which is generated by unresolved phase effects or the ultra-small-angle x-ray scattering from unresolved sample microstructures. A quantitative measure of this dark-field signal can be useful in revealing the microstructure size or material for medical diagnosis, security screening and materials science. Recently, we derived a new method to quantify the diffusive dark-field signal in terms of a scattering angle using a single-exposure grid-based approach. In this manuscript, we look at the problem of quantifying the sample microstructure size from this single-exposure dark-field signal. We do this by quantifying the diffusive dark-field signal produced by 5 different sizes of polystyrene microspheres, ranging from 1.0~{\textmu}m to 10.8~{\textmu}m, to investigate how the strength of the dark-field signal changes with the sample microstructure size, \(S\). We also explore the feasibility of performing single-exposure dark-field imaging with a simple equation for the optimal propagation distance given microstructure with a specific size and thickness, and successfully verify this equation with experimental data. Our theoretical model predicts that the dark-field scattering angle is inversely proportional to \(\sqrt{S}\), which is consistent with our experimental data. 
\end{abstract}
\begin{document}

\flushbottom
\maketitle
%
%
\thispagestyle{empty}

\section*{Introduction}

With advancements in x-ray generators and detectors, as well as the introduction of computed tomography \cite{hounsfield,ambrose1973computerized}, x-ray imaging has become a widely used technique that non-invasively reveals the internal structure of a sample. Conventional x-ray imaging manifests the difference in the attenuating ability of different materials in the sample, to create image contrast. It is now one of the standard imaging techniques used in clinical practice, materials science and security screening \cite{russo2017handbook}. However, the image contrast is significantly degraded for samples made of weakly attenuating materials, such as soft biological tissues. In recent decades, advanced x-ray imaging techniques known as phase contrast x-ray imaging (PCXI) have been developed, to enhance the image contrast for samples made up of materials that have similar attenuating properties. PCXI covers a set of techniques that convert the phase shift experienced by the x-ray wavefield while passing through the sample, into an intensity modulation that can be measured on a detector.  Examples include propagation-based imaging (PBI) \cite{snigirev1995possibilities,cloetens1996phase,wilkins1996phase,paganin2002simultaneous}, analyser-based imaging (ABI) \cite{davis1995phase,chapman1997diffraction,pagot2003method,bravin2003exploiting}, grating-interferometry (GI) \cite{david2002differential,momose2003demonstration,weitkamp2005x, pfeiffer2006phase}, edge-illumination (EI) \cite{olivo2001innovative,olivo2007coded,olivo2007modelling,munro2012phase}, single-grid imaging \cite{wen2010,morgan2011_grid}, and speckle-based imaging \cite{morgan2012,berujon2012_sb}.

The size of the smallest detectable sample feature, in an x-ray imaging system, is typically restricted by the spatial resolution of the system. Diffusive dark-field imaging (henceforth termed `dark-field imaging' for simplicity) is a way around this limit.  Such dark-field imaging looks at diffuse scattering---e.g.~small-angle x-ray scattering (SAXS) or ultra-small-angle x-ray scattering (USAXS)---from sample microstructures, in order to detect their presence. The dark-field signal is useful since it can reveal the presence of spatially random sample microstructure, which is otherwise invisible when using the full-field conventional or phase-contrast x-ray imaging techniques. This imaging modality is also more dose-efficient, since detectors with larger pixel sizes can be used than if the features were to be resolved directly. 

The dark-field signal has been measured qualitatively using most of the PCXI techniques, either via an approach where the dark-field information is extracted from contrast seen across a neighbourhood of pixels (such as PBI \cite{gureyev2020,leatham2021x} and single-grid imaging \cite{wen2010,morgan2013,croughan2022directional}), or via an approach where the dark-field is extracted on a pixel-by-pixel basis from multiple exposures (such as ABI \cite{rigon2007_abi_3}, GI \cite{pfeiffer2008}, and EI \cite{endrizzi2014ei,endrizzi2017x,matsunaga2019}). One interesting case is multiple-exposure speckle-tracking, where dark-field is extracted from local neighbourhoods of pixels across multiple exposures. It is worth noting that the dark-field signal has been retrieved from speckle-based set-ups using both explicit \cite{berujon2012_sb,zanette2014,zdora2017x} and implicit \cite{pavlov2020x,alloo2022dark, beltran2022} approaches to analyse how the speckles change. In explicit speckle-tracking approaches, the changes/motions in the speckles are tracked in each local neighbourhood of pixels, whereas in implicit speckle-tracking approaches, the changes in the speckles are tracked by looking at how the intensity translates and diffuses across the whole image, using one whole-image analysis step. The potential of dark-field imaging can be further explored by quantifying the dark-field signal, and then relating the signal strength to the properties of the sample microstructure, such as the size, material or the arrangement of the individual microstructures.  

There are various fields that can potentially benefit from the quantification of dark-field signal, including medical diagnosis, security screening and materials science. Some possible biomedical applications include imaging lungs with emphysema to measure airway size \cite{yaroshenko2013pulmonary,kitchen2020emphysema}, imaging breast tissues with microcalcifications for early detection of cancer \cite{michel2013dark, aminzadeh2022}, and imaging kidney stones of different compositions and microscopic morphology for classification \cite{scherer2015non}. Multiple animal studies have shown that lung diseases such as lung cancer \cite{scherer2017x}, emphysema \cite{hellbach2015vivo} and fibrosis \cite{hellbach2017x} can result in a weaker dark-field signal from the lungs, due to the change in size or structure of the alveoli. Recently, the diagnostic capability of a quantitative dark-field signal has also been demonstrated on healthy individuals \cite{gassert2021x} and chronic obstructive pulmonary disease (COPD) patients with emphysema \cite{willer2021x}, where the dark-field signal was correlated to the lung volume and the diffusion capacity of carbon monoxide, respectively. Other possible safety or industrial applications include imaging and/or detecting goods that come in powder form, such as drugs or explosives \cite{miller2013phase}, and imaging industrial parts made from carbon fibres \cite{valsecchi2020}.

A quantitative x-ray dark-field signal has been successfully extracted and related to the sample properties, using ABI \cite{kitchen2020emphysema}, GI \cite{prade2016,lynch2011_coeff,bech2010_coeff,gkoumas2016,harti2017} and EI \cite{modregger2017} techniques. The dark-field signal extracted from these techniques has been related to different sample microstructure properties, such as (i) the number of scattering interfaces, which can be related to the number of microstructures \cite{kitchen2020emphysema}, (ii) the kurtosis, which is a statistical quantity of the scattering distribution that can be related to the microstructure size \cite{modregger2017}, (iii) the correlation length, which is the length at which the correlation between the microstructures is probed by GI \cite{prade2016,harti2017}, and (iv) the linear diffusion coefficient (or dark-field extinction coefficient), which is analogous to the linear attenuation coefficient, and relates to the second statistical moment or width of the scattering probability distribution function of the sample \cite{bech2010_coeff,lynch2011_coeff,gkoumas2016}. However, the x-ray dark-field signal has not yet been quantitatively related to sample properties using the single-grid imaging technique. 

The single-grid imaging technique \cite{wen2010,morgan2011_grid} is a grating-based PCXI technique, which is similar to GI, but with a relatively simple setup compared to other dark-field imaging techniques, such as ABI, GI, and EI.  See Fig.~\ref{fig:sample_im}. As its name implies, the single-grid imaging technique only requires one optical element (a grid), and neither calibration nor alignment is required prior to the data acquisition. The detector needs to have a pixel size smaller than the grid period, so that the intensity pattern formed by the grid can be fully resolved. Both absorption grids \cite{wen2010,bennett2010_invivo, morgan2011_grid} and phase grids \cite{morgan2013,rizzi2013x, gustschin2021} can be used in this technique and the grid can be placed immediately upstream or downstream of the sample. The grid can be replaced by any object that provides an intensity pattern with high visibility, for example, a piece of sandpaper, in which case the technique is known as speckle-based imaging \cite{morgan2012,berujon2012_sb}. A grating can also be used in this technique, but it is less favourable than a grid, since the system will only be sensitive to the differential phase in one direction (perpendicular to the grating lines) \cite{morgan2011_grating}. The two-dimensional sensitivity provided by a grid is essential in the reconstruction of artefact-free projected phase images \cite{kottler2007}, which can be useful in quantifying microstructure when used in conjunction with dark-field images \cite{kitchen2020emphysema}. 

The data acquisition process of single-grid imaging involves only a single sample exposure, where the grid patterns the illumination. The sample-and-grid image is then compared to the reference image taken without the sample, where only the grid is present. This permits the simultaneous extraction of  attenuation, phase-shift and dark-field signals produced by the sample. These three quantities result in a decrease in mean, a shifting, and a broadening of the intensity pattern, respectively. The short data acquisition time can minimise motion blurring and x-ray radiation dose, which makes this technique feasible for dynamic imaging.  

Recently, a new retrieval algorithm by How \& Morgan \cite{how2022quantifying} was derived to quantify the x-ray dark-field signal in single-grid imaging and relate the signal to the number of microstructures, \(N\). The algorithm was applied to a sample with unresolved microstructures, made up of 1.0~{\textmu}m polystyrene microspheres. Below, we apply the same algorithm to samples made up of 5 different sizes of polystyrene microspheres, all smaller than the resolution of the imaging system.  This allows us to investigate how the strength of the dark-field signal changes with the sample microstructure size, and determine the feasibility of performing single-exposure quantitative dark-field imaging using the single-grid imaging technique.

We first provide a mathematical model relating how the effective scattering angle extracted is related to the sample microstructure size. This is achieved by relating the number of microstructures in the beam path of an x-ray passing through the sample to the total number of microstructures in the sample, which is then related to the total volume of the microstructures and subsequently the total thickness of the microstructures. We show that the experimental data are consistent with this model, by plotting the scattering angle as a function of sample thickness rather than sample microstructure size. We then derive an expression for the propagation distance at which to perform single-exposure quantitative dark-field imaging with maximum sensitivity. This is achieved by analytically solving for the propagation distance at which the change in dark-field signal with respect to the change in scattering angle is maximised. Via this expression, we compute a range of suitable sample-to-detector distances for each sample microstructure size, and we verify this by comparing (i) the effective scattering angle extracted using a single exposure, to (ii) the effective scattering angle extracted from multiple distances. Finally, we explore the effects of propagation-based phase contrast effects overlying the modelled grid and dark-field effects, discuss the properties of this technique, and look at potential applications as well as future research directions.          

\section*{Mathematical modelling}

Here, we apply the algorithm developed by How \& Morgan, to quantify the dark-field signal in single-grid imaging \cite{how2022quantifying}. In this approach, the dark-field signal is extracted using an explicit cross-correlation approach between the grid-only image and the grid-and-sample image. A summary of the algorithm is given below, with full details available in How \& Morgan \cite{how2022quantifying}. 

\subsection*{Extraction of the quantitative dark-field signal using an explicit cross-correlation approach}
The x-ray intensities seen at the detector in the presence of the grid, \(I_{g}(x)\), and the grid and sample, \(I_{sg}(x)\), are defined as sine functions in one-dimension, which are given by 
\begin{equation}
I_{g}(x) = a \sin \left(\frac{2 \pi  x}{p}\right)+b
\end{equation}
and 
\begin{equation}\label{Isg}
    I_{sg}(x) = A a \sin\left(\frac{2 \pi  x}{p}\right) + t b,
\end{equation}
respectively.  Here, $x$ is the position across the sample, \(a\) is the amplitude, \(b\) is the mean, and \(p\) is the period of the intensity oscillations due to the grid, \(A\) is the change in amplitude of the grid intensity oscillations that is introduced by the sample, and \(t\) is the transmission of the x-ray wavefield passing through the sample. The dark-field signal, \(DF\), which is defined as the relative change in visibility between the sample-grid intensity (or stepping curve), \(V^{s}\), and the grid-intensity (or stepping curve), \(V^{r}\) for a grating-based method \cite{pfeiffer2008}, is  
\begin{equation}\label{DF_signal}
\begin{split}
    DF &= \frac{V^{s}}{V^{r}} = \frac{\frac{Aa}{tb}}{\frac{a}{b}} 
    = \frac{A}{t}.
\end{split}
\end{equation} 
Here, the values of \(A\) and \(t\) are determined by fitting the local cross-correlation results of the grid image \(I_{g}(x)\) both with itself, and with the sample and grid image \(I_{sg}(x)\). The dark-field visibility signal has a value between \(0\) and \(1\), where \(1\) represents no dark-field signal/scattering and \(0\) represents maximum dark-field signal, where the grid is invisible and the cross-correlation (or stepping curve) is `flat'.   

By modelling the blurring kernel applied to the grid pattern in the presence of the sample as a normalised zero-centred Gaussian function  \cite[Figure~5]{ kitchen2004}\cite{khromova2004monte,khelashvili2005_gauss}, and \(I_{sg}(x)\) as the convolution between \(I_{g}(x)\) and the Gaussian function, the dark-field signal is 
\begin{equation}\label{df_dist_gauss}
    DF = \frac{V^{s}}{V^{r}} = \frac{a \exp\left(-\frac{2 \pi ^2 (d/2)^2}{p^2}\right)}{b} \frac{b}{a} = \exp \left(-\frac{2 \pi ^2 (d/2)^2}{p^2}\right) = \exp \left(-\frac{ \pi ^{2} z^{2} \theta^{2}}{2p^2}\right).
\end{equation}
Above, we used the scattering width, \(d = z \theta\), where \(z\) is the sample-to-detector propagation distance and \(\theta\) is the effective scattering angle. Equation~(\ref{df_dist_gauss}) describes how the dark-field signal changes with propagation distance, which can be fitted to dark-field signals measured at one or many different propagation distances, to accurately extract the effective scattering angle.   

The effective scattering angle can be related to the number of microstructures, \(N\), in the paraxial ray path, a number which is proportional to the total sample thickness, \(T\), if we assume these  microstructures---which in our experiment are microspheres---to have the same size. It has been observed, using both crystal-analyser-based imaging and single-grid imaging, that the x-ray scattering angle is proportional to \(N\) to a power that is greater than $\tfrac{1}{2}$ \cite{kitchen2020emphysema,how2022quantifying}. This deviates from the random-walk model proposed by von Nardroff \cite{von1926refraction}, in which the scattering angle is proportional to \(\sqrt{N}\). The relationship between the scattering angle and the number of microstructures can be written as
\begin{equation} \label{theta_and_n}
    \theta = k N^{\frac{1}{2} + \alpha}, 
\end{equation}
where \(k\) is a positive coefficient and \(\alpha\) is an anomalous diffusion constant\cite{MetzlerKlafter2000} that can be greater than or smaller than 0.  

\subsection*{Relating the scattering angle to the sample microstructure size}
The algorithm proposed in How \& Morgan \cite{how2022quantifying} has only been applied to a sample with microstructure size of 1.0~{\textmu}m.  Thus it is interesting to apply the algorithm to samples that have different microstructure sizes, to investigate how the strength of the dark-field signal changes with microstructure size. 

We start by deriving the relationship between \(N\) and the microsphere size. First, assume a rectangular cuboid with x-rays normally incident on an area, \(A_{c}\), of one cuboid face.  The x-rays pass through the thickness \(T_{c}\), which is randomly filled with microspheres of diameter \(S\). The average number of microspheres, $N$, along the ray path, is the ratio between (i) the total projected line-of-sight area of the spheres in the cuboid, and (ii) the area of the entrance face of the cuboid. Hence 
\begin{equation} \label{n_and_d}
    N = \frac {n_{T} \pi (S/2)^{2}}{A_{c}}  
    = \frac{V_{T}}{\frac{4}{3}\pi (S/2)^{3}} \frac{\pi (S/2)^{2}}{A_{c}} 
    = \frac{A_{c}T}{\frac{4}{3}\pi (S/2)^{3}} \frac{\pi (S/2)^{2}}{A_{c}} 
    = \frac{3T}{2S}, 
\end{equation}
where \(n_{T}\) is the total number of microspheres in the cuboid and \(V_{T}\) is the total volume of microspheres. Note that, after the third equals sign, we replaced \(V_{T}\) with \(A_{c} T\). We can imagine this volume as melting all microspheres into a rectangular cuboid with area \(A_{c}\) and width \(T\), which is essentially the thickness of microspheres in the beam path of an x-ray wavefield passing through the cuboid (as would be measured in an attenuation signal). 

Since we do not expect the anomalous diffusion coefficient \(\alpha\) to change for microspheres of different sizes (as shown by Kitchen \textit{et al}. \cite{kitchen2020emphysema}), we assume \(\alpha\) to be \(0\), which is consistent with the model of von Nardroff\cite{von1926refraction}, to compare the strength of the dark-field signal from microstructures of different sizes via the \(k\) value in Eqn.~(\ref{theta_and_n}). Using Eqn.~(\ref{n_and_d}), the $\alpha=0$ case of Eqn.~(\ref{theta_and_n}) gives
\begin{equation}\label{theta_and_D}
    \theta = k \sqrt{N} = k \sqrt{\frac{3T}{2S}} = K \sqrt{T},
\end{equation}
where \(K = k \sqrt{3/(2S)}\).  We see that \(\theta\) is proportional to $1/\sqrt{S}$. 

\subsection*{Optimal distance for single-exposure dark-field imaging}
\label{s:opt distance for single exp}  
\begin{figure}[hbt!]
\centering
\includegraphics[width=0.45\linewidth]{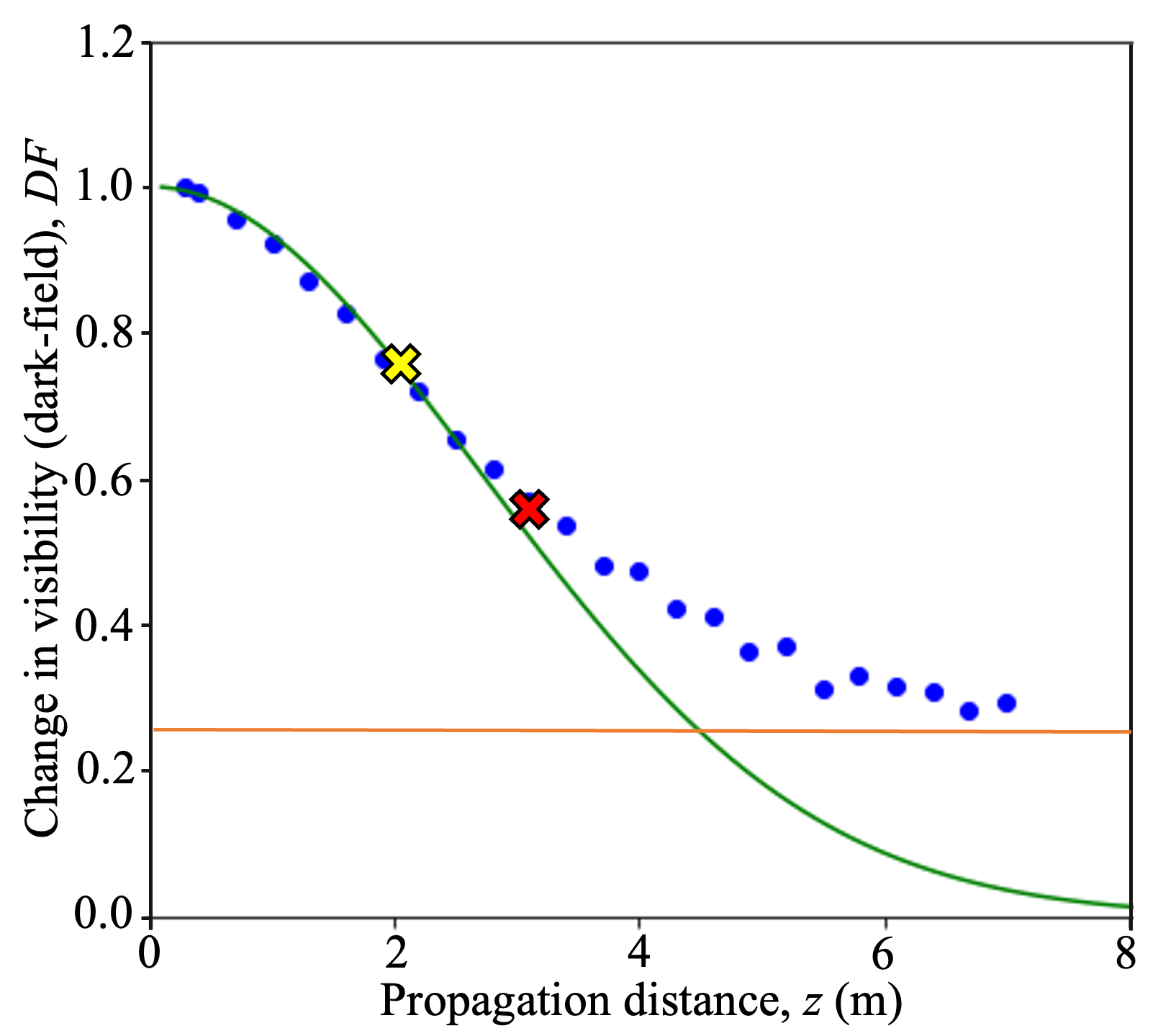}
\caption{A typical example of the dark-field signal measured as a loss in visibility (at pixel (960, 1001), from the 8.0~{\textmu}m sample discussed later) from 24 propagation distances. The dark-field signal starts to saturate at a propagation distance of around 3.1 m (labelled with the red cross) and thus the dark-field signals measured at 3.1 m and beyond are not included in the fitting to extract the effective scattering angle using Eqn.~(\ref{df_dist_gauss}). The yellow cross labels the maximum-gradient point where the change in visibility with respect to the change in scattering angle/propagation distance is maximised. This specifies where the sensitivity of the dark-field imaging system is maximised, providing the optimal distance to perform single-exposure quantitative dark-field imaging. This optimal distance for sample microstructure of different sizes is given by Eqn.~(\ref{maths_opt_distance}). The orange line indicates the `visibility floor', at which the decrease in visibility is dominated by the source-size blurring effect on the reference grid pattern, and the signal retrieved is not representative of the sample anymore.}
\label{fig:DF_dist}
\end{figure}

The optimal distance for single-exposure dark-field imaging is the distance at which the sensitivity of the imaging system towards the dark-field signal is maximised. This means that a change in the scattering angle produced by the sample (e.g.~due to a different microstructure size) results in the biggest possible change in the measured dark-field visibility signal (see the yellow cross in Fig.~\ref{fig:DF_dist}). This can be determined  by analytically solving for the propagation distance, \(z_{opt}\), for which  \(\frac{\partial^{2} (DF)}{\partial \theta ^{2}} = 0\), where \(DF\) is defined in Eqn.~(\ref{df_dist_gauss}). Hence    
\begin{equation} \label{maths_opt_distance}
    z_{opt} = \frac{p}{\pi \theta} = \frac{p}{\pi K \sqrt{T}} = \frac{p}{\pi \left(\frac{1}{\sqrt{mS}}\right)\sqrt{T}} = \frac{p}{\pi}\sqrt{\frac{mS}{T}}.
\end{equation}
Note, after the second equals sign, we used the expression for \(\theta\) in  Eqn.~(\ref{theta_and_D}). Furthermore, after the third equals sign, we substituted in \(K = 1/\sqrt{mS}\), since by the definition of \(K\) just below Eqn.~(\ref{theta_and_D}),
\begin{equation}
    \frac{1}{K^{2}} = \left(\frac{2}{3k^{2}}\right) S = mS,
\end{equation}
where $m=2/(3k^2)$ is the gradient of a plot of $1/K^2$ against microstructure size \(S\) (see Fig.~\ref{fig:coeff_against_diameter} (b)).

\subsection*{Saturation of dark-field signal}
\label{s:DF_sat_def}
Saturation of the dark-field signal can be observed at overly large propagation distances. This introduces a limit to where the equations described above are physically useful.  Beyond this limit the grid intensity pattern becomes too blurred, with the propagation distance being so large that the algorithm can no longer recognise the reference pattern and thus fails to properly fit the cross-correlation curves. In grating interferometry, the sensitivity of the imaging system towards the dark-field signal is typically tuned accordingly, to make sure that the dark-field visibility signal is maintained above a given value, e.g. 0.1. In other words, the reduction in visibility of the stepping curve is kept below 90$\%$ \cite{gradl2018dynamic}, to avoid obtaining a saturated dark-field signal. Similarly, in the single-grid imaging technique, we can tune the sensitivity to make sure the dark-field signal is maintained above 0.3 or 0.4 (as suggested by our data). We use a higher visibility threshold because the signal is extracted from a single exposure, which is more susceptible to noise compared to the signal extracted from the stepping curve obtained in GI using multiple exposures. Also, our technique requires the visibility of the grid to be stronger than the visibility of surrounding sample features, which is not the case in GI. This tuning can be achieved by taking the images at a suitable propagation distance. In this manuscript, when fitting across multiple distances we have excluded the dark-field signal measured beyond a certain threshold distance---where the dark-field signals begin to saturate (as shown in Fig.~\ref{fig:DF_dist})---during the extraction of the effective scattering angle. This threshold distance is determined by looking at how the dark-field signal from samples of different thicknesses changes with propagation distance. Note, the threshold distance for each sample is different (as shown in the \hyperref[s:num_analysis]{Numerical analysis} section) since the number of interfaces encountered by the x-ray beam is different for samples of the same projected thickness but different microstructure sizes.     

The dark-field signal may also become saturated due to additional visibility contributed by the speckle pattern formed by the microspheres or other surrounding sample features. It has been demonstrated, via experiment and simulation, that a speckle pattern can be formed by randomly-packed glass microspheres due to multiple-beam refraction and free-space propagation \cite{kitchen2004}. The intensity variations that make up the speckle pattern may locally enhance the visibility of the observed reference intensity pattern and thus result in an `increase' of the dark-field visibility signal. A detailed simulation \cite{kitchen2004} is required to determine the contribution of the speckle pattern formed by the microspheres to the dark-field signal, which is outside of the scope of this manuscript.

\section*{Methods}
We captured an experimental dataset of the sample shown in Fig.~\ref{fig:sample_im}, which includes different microstructure sizes and a range of projected thicknesses. We imaged over a range of propagation distances, to investigate (i) how the dark-field signal changes with the size of microstructure, and (ii) at which distance quantitative single-exposure imaging is optimum. 

\subsection*{Experimental setup}
Our experimental setup is shown in Fig.~\ref{fig:sample_im}.  This is a typical single-grid imaging setup, with the addition that the propagation distance is allowed to change. The experiment was performed at the Australian Synchrotron Imaging and Medical Beamline (IMBL). An attenuating grid (a geological stainless steel sieve, as used in How \& Morgan \cite{how2022quantifying}) was placed 43~cm upstream of the sample (i.e.~as close as possible). The sample and grid were placed on two different stages which could be moved in a horizontal direction automatically, so that they could be moved out of the field of view to capture grid-only and flat-field images. A 25~{\textmu}m thick Gadox phosphor was coupled to a pco.edge 5.5 sCMOS detector to collect images and it was placed on a separate table which could be moved to adjust the propagation distance $z$ between the sample and the detector. The energy of the x-rays was 34 keV and the effective pixel size of the setup was 9.8~{\textmu}m.  

\subsection*{Sample preparation}
\begin{figure}[htbp]
\centering
\includegraphics[width=0.8\linewidth]{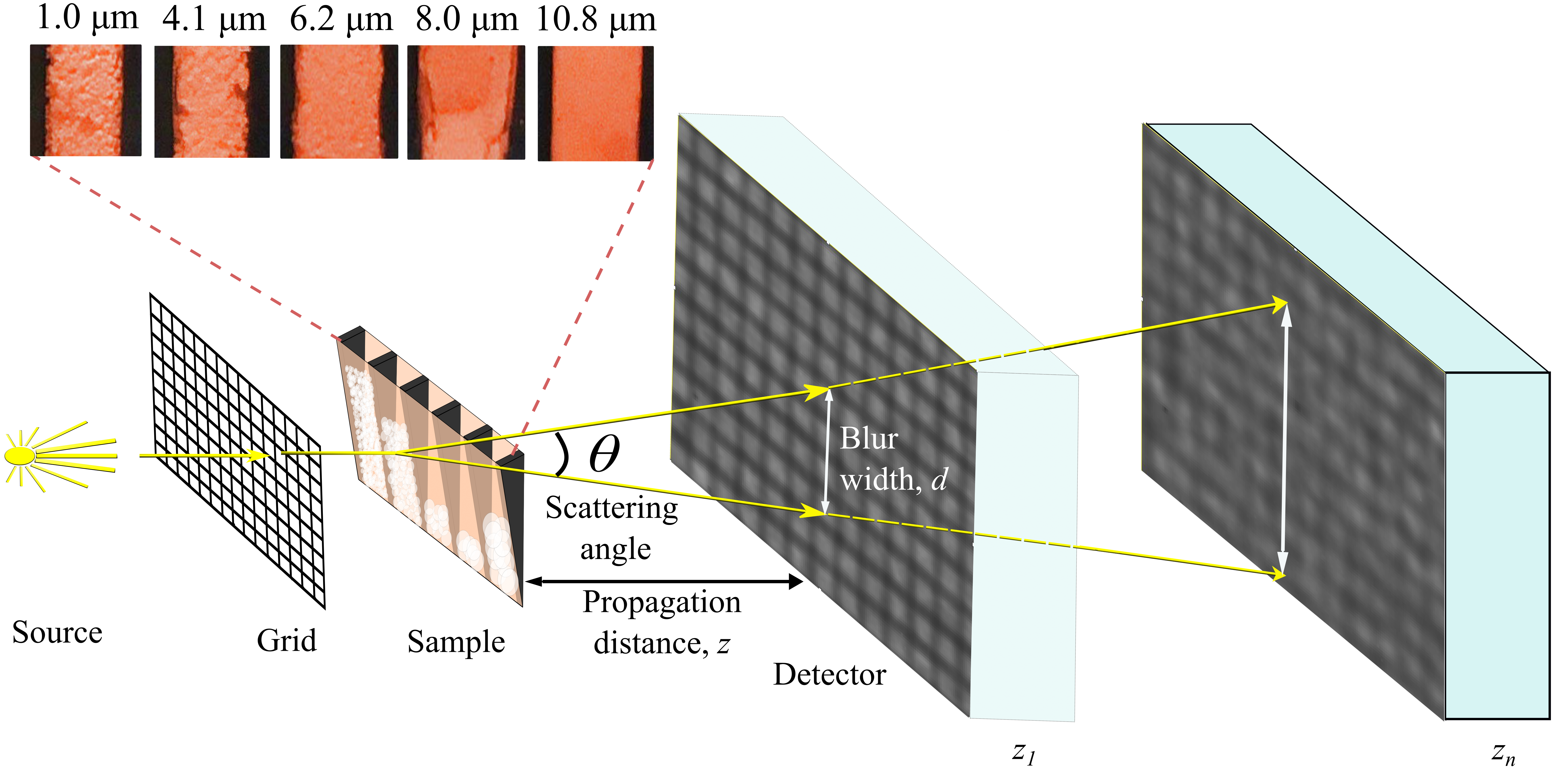}
\caption{Experimental setup for single-grid imaging with the detector placed at multiple propagation distances, \(z_n\). Images are taken with and without the sample. Polystyrene microspheres of diameter 1.0~{\textmu}m, 4.1~{\textmu}m, 6.2~{\textmu}m, 8.0~{\textmu}m and 10.8~{\textmu}m (from left to right, viewed from the source) were placed in a custom-made sample holder made up of a Kapton sheet and rubber wedges. The inset shows the zoomed-in view of each sample `cell' through the orange Kapton. The blur width increases as the propagation distance increases, resulting in the grid intensity pattern being smeared out more significantly and producing a stronger dark-field signal. Note that the scattering angle $\theta$ has been exaggerated for visualisation purposes.}
\label{fig:sample_im}
\end{figure}

The polystyrene microspheres were purchased suspended in 10~ml of water (Microspheres-Nanospheres, Corpuscular Inc., Cold Spring New York, USA), with the spheres making up 2.5\% of the volume. Sample tubes were placed into a centrifuge to separate the microspheres and water. The water was removed using a pipette and the cap was left off to allow any remaining water to evaporate. The sample tubes were placed in a water bath sonicator to break up clumps formed during the liquid evaporation process.  Nevertheless, there were obvious clumps remaining, as seen in the inset of Fig.~\ref{fig:sample_im}. The microspheres were then transferred into the sample holder. 

A sample holder with five `cells' was custom-made with a piece of Kapton sheet, which provides minimal attenuation to x-rays, with each `cell' separated by rubber wedges, glued to the Kapton, to provide a range of sample thicknesses as shown in Fig.~\ref{fig:sample_im}. The five `cells' held microspheres of diameter 1.0~{\textmu}m, 4.1~{\textmu}m, 6.2~{\textmu}m, 8.0~{\textmu}m and 10.8~{\textmu}m.

\subsection*{Data acquisition}
Flat-field (without grid or sample), grid-only and sample-grid images were taken at sample-to-detector propagation distances of 0.28 m, then 0.4 m to 7 m at steps of 0.3 m. One set of dark-current images was taken at the end of the experiment. The exposure time was chosen to be 160 ms to fill the dynamic range of the detector and 30 exposures were taken for each set of images, which were averaged together prior to analysis to reduce the noise level. Because the sample was wider than the field of view, two neighbouring sample-grid images were taken at each distance (with the 6.2~{\textmu}m sample appearing in both the left and right images). The images shown in this manuscript place the two sample images immediately next to each other.

\subsection*{Numerical analysis}
\label{s:num_analysis}
The raw images were first flat and dark-corrected, demagnified to account for the subtle magnification effect seen especially at larger propagation distances, and cropped to reduce the processing time. The images taken at 24 propagation distances were analysed using the algorithm described in How \& Morgan \cite{how2022quantifying}, with a cross-correlation window size of 14 pixels to match the grid period. The dark-field signals measured at 24 distances were then fitted to Eqn.~(\ref{df_dist_gauss}) to extract the effective scattering angle, \(\theta\). The dark-field images taken at different propagation distances were realigned using the linear alignment function in ImageJ software that applies the Scale Invariant Feature Transform (SIFT). The image transformation matrices used to realign the dark-field images were obtained from the alignment of the transmission images. The dark-field images were also smoothed by a median square kernel of size 14 pixels (i.e.~the grid period) before the extraction of the scattering angle, to reduce the noise level. As mentioned in the \hyperref[s:DF_sat_def]{Saturation of dark-field signal} section, a threshold distance was determined for each sample to avoid saturation of the dark-field signal, which was 1.6 m for the 1.0~{\textmu}m microspheres, 2.5 m for the 4.1~{\textmu}m microspheres and 3.1 m for the remaining samples. The dark-field signals measured at the threshold distance and beyond are not included in the fitting to extract the effective scattering angle, as shown in Fig.~\ref{fig:DF_dist}.

\section*{Results}
\subsection*{Quantifying microstructure size from the x-ray dark-field signal}
\begin{figure}[hbt!]
\centering
\includegraphics[width=0.8\linewidth]{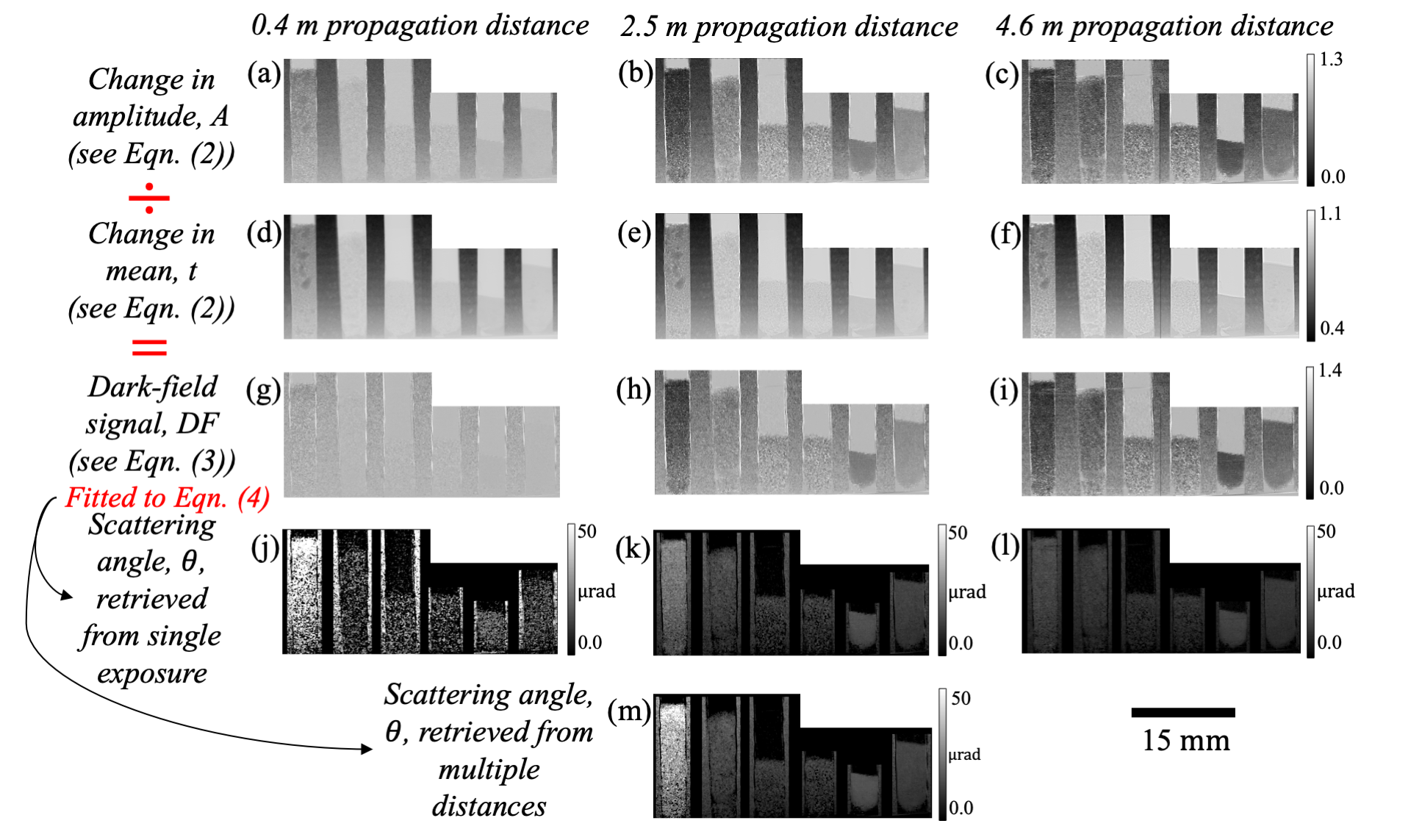}
\caption{Dark-field imaging results obtained from single-grid images of polystyrene microspheres of 5 different sizes (with the 6.2~{\textmu}m sample appearing in both the left and right images of the image pair), separated by rubber wedges, shown here for three different propagation distances -- 0.4 m, 2.5 m and 4.6 m. The diameter of the microspheres in each panel (from left to right, similarly for the rest of the figures in this paper unless specified otherwise) is 1.0~{\textmu}m, 4.1~{\textmu}m, 6.2~{\textmu}m, 6.2~{\textmu}m, 8.0~{\textmu}m and 10.8~{\textmu}m. The change in amplitude of intensity oscillations, \(A\) ((\textbf{a}), (\textbf{b}) \& (\textbf{c})), is divided by the corresponding transmission of the x-ray wavefield, \(t\) ((\textbf{d}), (\textbf{e}) \& (\textbf{f})) to obtain the dark-field signal, \(DF\) ((\textbf{g}), (\textbf{h}) \& (\textbf{i})). The effective scattering angles, \(\theta\), shown in panels (\textbf{j}), (\textbf{k}) and (\textbf{l}) are extracted from the \(DF\) signal in panels (g), (h) and (i) respectively, using Eqn.~(\ref{df_dist_gauss}), while \(\theta\) in panel (\textbf{m}) is extracted from the \(DF\) images taken at 24 distances, using Eqn.~(\ref{df_dist_gauss}). The microspheres with a larger size produce a weaker dark-field signal and thus a smaller scattering angle, which agrees with Eqn.~(\ref{theta_and_D}). It is difficult to see this when comparing the 8.0~{\textmu}m and 10.8~{\textmu}m samples to the others, since these microspheres are more densely packed. The \(\theta\) image extracted from the larger propagation distance has less noise than the smaller propagation distance, where the blurring effect is weak. The scattering angle extracted from a shorter propagation distance ((j)) is larger compared to those extracted from a larger or multiple distances. The scattering angle extracted from a propagation distance of 2.5 m was consistent with the scattering angle extracted from multiple distances (except the 1.0~{\textmu}m sample), indicating that this is a suitable propagation distance for single-exposure dark-field imaging of sample microstructure of these sizes.}
\label{fig:combine}
\end{figure}

Figure~\ref{fig:combine} shows the results obtained from the images taken at propagation distances of 0.4 m, 2.5 m and 4.6 m, including the change in amplitude (\(A\)), transmission (\(t\)), dark-field signal (\(DF\)) and the effective scattering angle (\(\theta\)) extracted using the dark-field signals captured from single and multiple distances. The sample with the smallest microstructure, 1.0~{\textmu}m (first `cell' from the left), produces the strongest dark-field signal and thus the largest effective scattering angle.  The strength of the dark-field signal decreases with the sample microstructure size, which is consistent with the inverse square root relationship between the \(\theta\) and microstructure size (\(S\)) in Eqn.~(\ref{theta_and_D}). This is because the x-rays are being scattered by more interfaces as they pass through a sample with a smaller microstructure size, compared to the sample of the same thickness but with a larger microstructure size. It is worth noting that the dark-field signals produced by the 8.0~{\textmu}m and 10.8~{\textmu}m samples are stronger compared to the 6.2~{\textmu}m sample. This is due to the more effective packing of the microspheres in those two `cells', seen by fewer `clumps', and resulting in fewer air gaps, more microstructures and greater sample thickness than the 6.2~{\textmu}m sample.   

\begin{figure}[hbt!]
\centering
\includegraphics[width=0.8\linewidth]{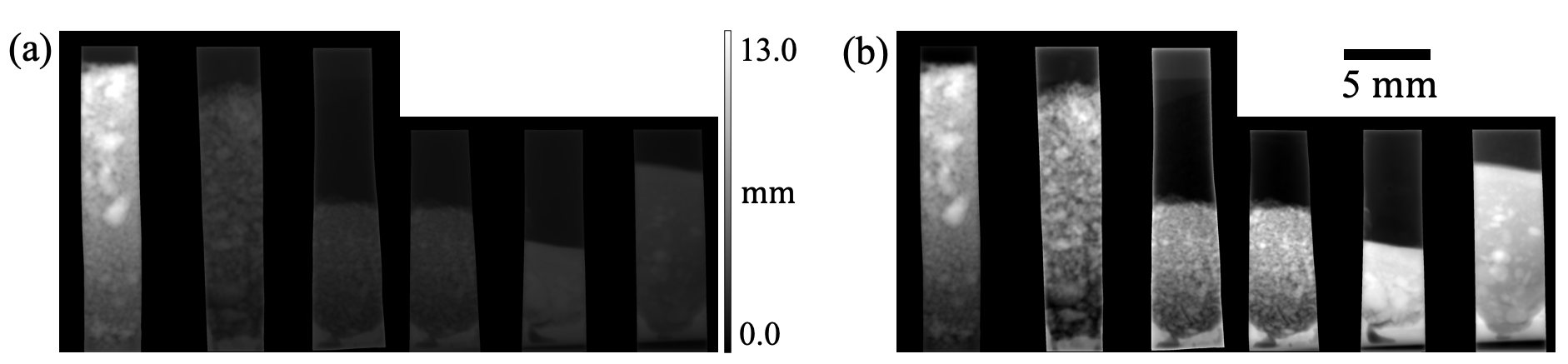}
\caption{Sample thickness, averaged from the thicknesses retrieved using the transmission images obtained at 1.9 m, 2.2 m, 2.5 m and 2.8 m, via the TIE-based single-material phase retrieval algorithm \cite{paganin2002simultaneous}. (\textbf{a}) Sample thickness image with the same greyscale for the whole image. (\textbf{b}) Sample thickness image with grey level of each sample `cell' normalised between \(0\) and \(1\) for visualisation purposes. Clumps with greater thicknesses shown in the figure were consistent with the clusters of microspheres observed by visual inspection of the sample (see Fig.~\ref{fig:sample_im}). The shape of the rubber wedges provides a wide range of projected sample thickness that increases gradually from the bottom to the top of the image. The 1.0~{\textmu}m and 8.0~{\textmu}m samples reach greater thicknesses compared to the other samples, due to the more effective packing of the microspheres.} 
\label{fig:thickness_mean_1928}
\end{figure}

\begin{figure}[hbt!]
\centering
\includegraphics[width=0.8\linewidth]{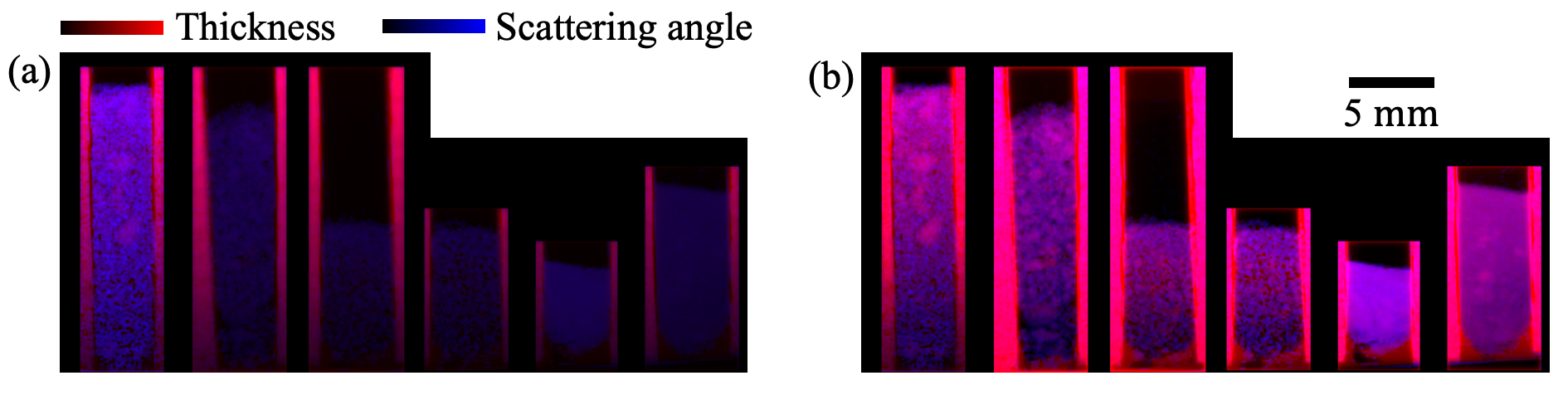}
\caption{Colour image of the samples that demonstrates the complementarity of the attenuation and dark-field signals. (\textbf{a}) Colour image with the same colour scale in the thickness and scattering angle image for all samples. (\textbf{b}) Colour image with the colour scale of thickness and scattering angle images normalised between 0 and 1 separately for each sample `cell'. The red and blue channels of the image correspond to the sample phase-retrieved thickness and scattering angle respectively. The rubber wedges have a stronger red hue but a weaker blue hue compared to the microspheres since they are more attenuating to x-rays, and contain fewer dark-field-producing unresolved microstructures.} 
\label{fig:colour_image}
\end{figure}

To study the dependence of the dark-field signal on microstructure size, we want an accurate measure of sample thickness, based on all collected data. The sample thickness (shown in Fig.~\ref{fig:thickness_mean_1928}) is obtained using a Transport of Intensity Equation (TIE)-based single-material phase retrieval algorithm \cite{paganin2002simultaneous} on the transmission image (e.g.~Fig.~\ref{fig:combine} (d)-(f)), and averaging the thickness retrieved from images collected at 1.9 m, 2.2 m, 2.5 m and 2.8 m. Microsphere clumps are clearly seen in both the photographs of the sample tubes in Fig.~\ref{fig:sample_im}, and the x-ray thickness image in Fig.~\ref{fig:thickness_mean_1928}. Since the rubber dividers were significantly more attenuating than the microspheres, they were excluded and those parts of the image were set to the average attenuation value, prior to the thickness retrieval process. This was done to obtain an accurate measure of the sample thickness from the single-material TIE algorithm, and avoid smoothing the contrast from the rubber into the region of the image containing microspheres, seen in Fig.~\ref{fig:colour_image}, where the same cropping is not used. The 1.0~{\textmu}m sample reaches a larger thickness compared to the other samples since the air-gaps between small microspheres are typically smaller. In Fig.~\ref{fig:thickness_mean_1928} (b), the contrast across all parts of the sample is adjusted by normalising the thickness of each `cell' between \(0\) and \(1\), to better visualise the details of the sample.  

Figure~\ref{fig:colour_image} shows the complementarity between the attenuation signal, shown in red, and the dark-field signal, shown in  blue. The microspheres produce a stronger dark-field signal relative to the attenuation signal when compared to the rubber wedges, and thus the microspheres are shown in a stronger blue hue than the rubber wedges. It is worth noting that certain regions in the 1.0~{\textmu}m and 10.8~{\textmu}m microsphere `cells' appear to have a slightly stronger red hue compared to the surrounding regions, as a result of the microspheres being packed more tightly together, potentially with some liquid remaining that reduces the relative strength of the dark-field signal compared to the attenuation signal. This agrees with what we observed in Fig.~\ref{fig:thickness_mean_1928}. In Fig.~\ref{fig:colour_image} (b), vertical red stripes (with no blue contribution) were observed near the inner edges of all rubber wedges surrounding the microspheres, suggesting maximum attenuation but minimum dark-field signal in these regions. We believe this is an artefact from the phase retrieval algorithm\cite{paganin2002simultaneous} as a result of our assumption that the whole sample is made up of a single material, polystyrene. Since the rubber wedges are more attenuating than the polystyrene, the phase retrieval algorithm ended up oversmoothing the phase near these edges and thus resulting in `extra' attenuation, interpreted as thickness, along the inner edge of each `cell'.      

\begin{figure}[hbt!]
\centering
\includegraphics[width=0.8\linewidth]{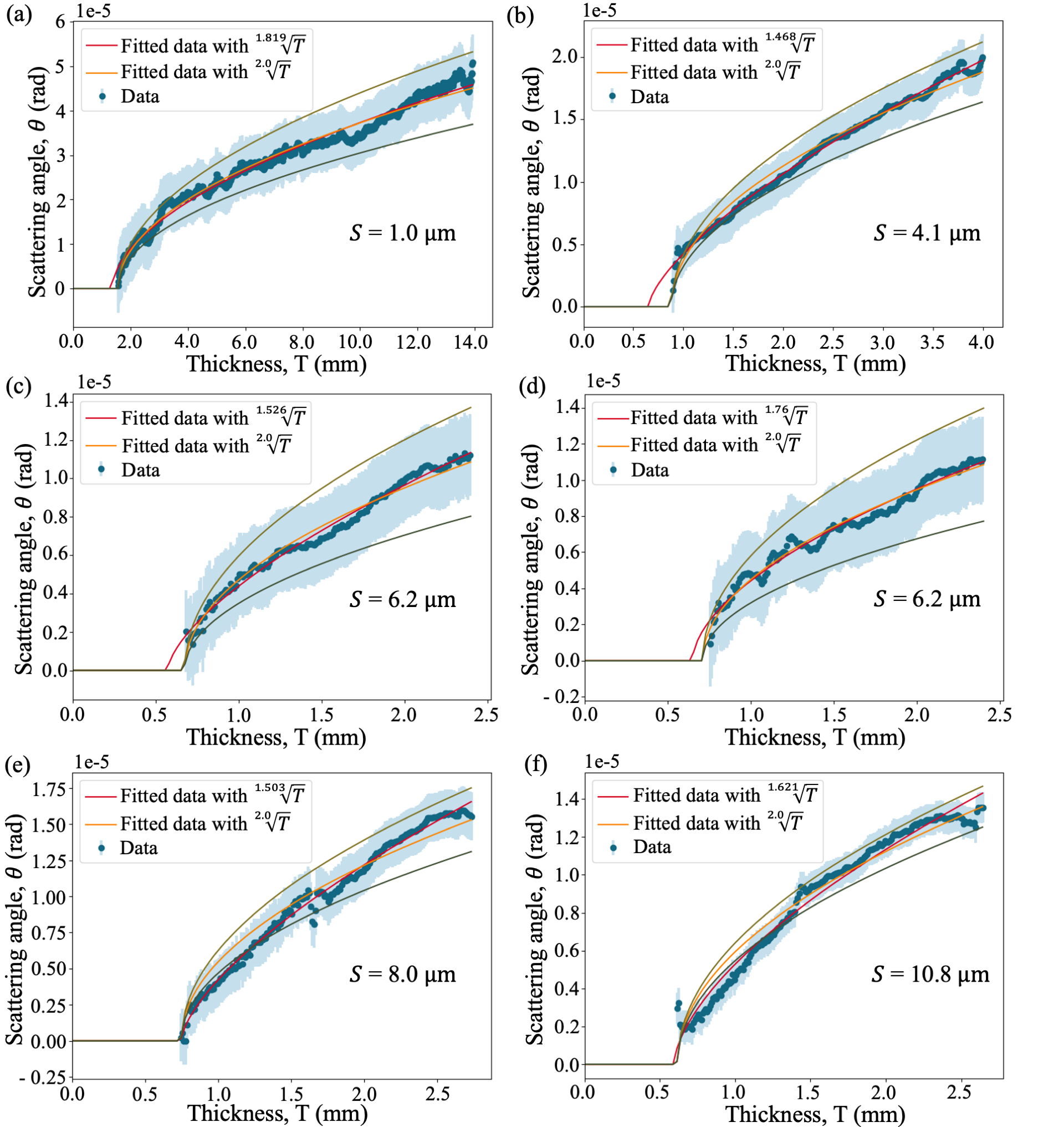}
\caption{Scattering angle, \(\theta\), extracted from microspheres of different diameters, $S$, as a function of sample thickness, \(T\), which is proportional to the number of microstructures, $N$, along a paraxial ray. Each plot is fitted with both \(\theta = K'\sqrt[p]{T-b'}\), where \(K'\) is a constant and the exponent is allowed to vary (red curves), and with \(\theta = K \sqrt{T-b}\), where \(K\) is a constant, as suggested by von Nardroff \cite{von1926refraction} (orange curves). The $x$-intercept, \(b'\) and \(b\), which represent the thickness of the Kapton sheet, is another variable in the fitting that can change. The uncertainty (shown in blue) is constant for each data point to make sure all data points have the same weighting during the fitting. The red curves provide a better fit to the data, suggesting the presence of anomalous diffusion. However, we focus on the orange curves since we are more interested in how the strength of the dark-field signal changes with sample microstructure size. The coefficient \(K\) is then related to the sample microstructure size as shown in Fig.~\ref{fig:coeff_against_diameter}.}
\label{fig:theta_thickness_fitted}
\end{figure}

We have now retrieved a quantitative measure of the scattering angle, which depends on both the sample and the x-ray illumination, so the next step is to relate the angle to the sample microstructure properties, such as the number of microstructures. The effective scattering angle extracted from each sample is then related to the sample thickness, which is proportional to the number of microstructures since we have microspheres of the same size. This is achieved by performing a least-squares fitting on the effective scattering angle as a function of sample thickness. We noticed that the least-squares fitting was greatly affected by the spread of data points. For example, the few data points describing smaller sample thicknesses would have less influence on the fit than the many data points at larger thicknesses since we have significantly fewer pixels measuring the smaller thicknesses. We overcame this by binning the data points based on their thickness and plotting the mean scattering angle value in each bin instead of plotting the values extracted from every pixel individually. The resulting plots are shown in Fig.~\ref{fig:theta_thickness_fitted}. The uncertainty of each data point is set to be the same, which is the median of the standard deviation of the angles obtained from all the bins in the `cell'. 

\begin{figure}[hbt!]
\centering
\includegraphics[width=0.8\linewidth]{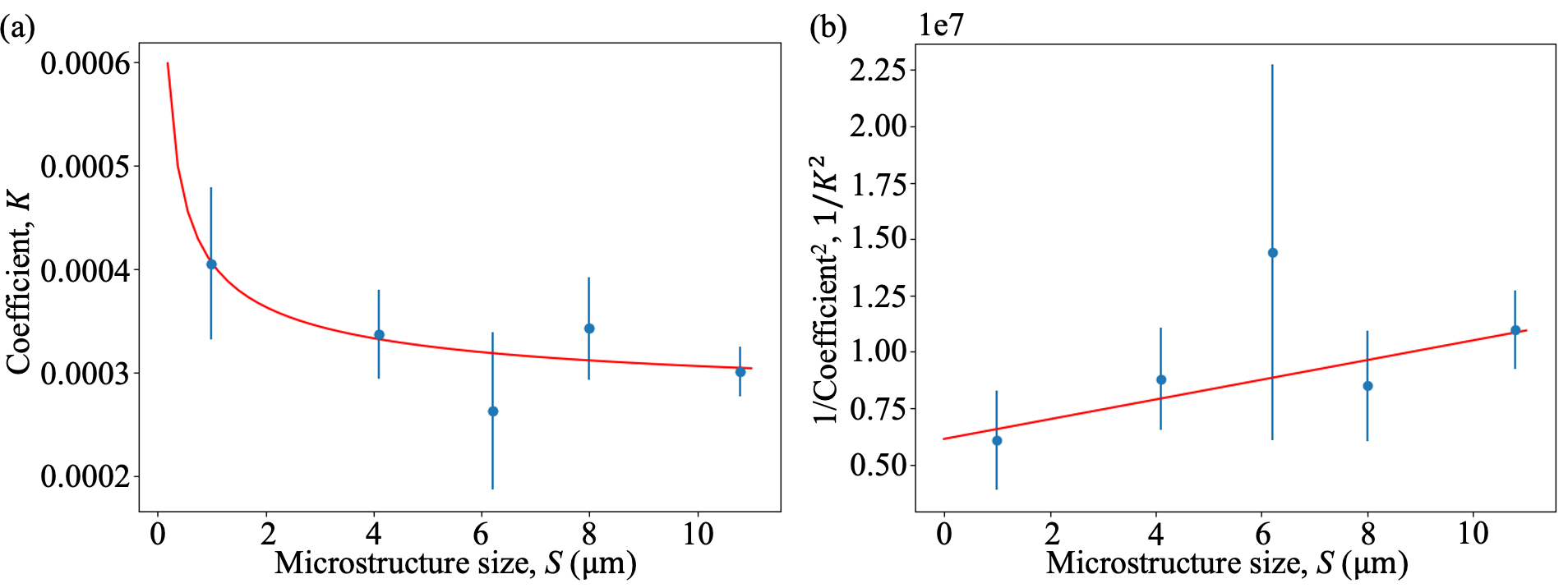}
\caption{The relationship between the fitted coefficient, \(K\) (obtained from Fig.~\ref{fig:theta_thickness_fitted}), and the sample microstructure size, \(S\), (\textbf{a}) before and (\textbf{b}) after linearisation. The results are consistent with a model that says $K$ is inversely proportional to \(\sqrt{S}\), which agrees with Eqn.~(\ref{theta_and_D}). Note that the data for 6.2~{\textmu}m shown here is extracted from Fig.~\ref{fig:theta_thickness_fitted} (d) and the uncertainty in each coefficient value is obtained by fitting two new functions to the data \(\pm\) uncertainties (green curves) in Fig.~\ref{fig:theta_thickness_fitted}. The gradient of the line of best fit in (b) was then used to solve for the optimal distance to perform single-exposure dark-field imaging, as explained in the \hyperref[s:opt distance for single exp]{Optimal distance for single-exposure dark-field imaging} section of the discussion.}
\label{fig:coeff_against_diameter}
\end{figure}

The Fig.~\ref{fig:theta_thickness_fitted} visibility plots are each fitted with both \(\theta = K'\sqrt[p]{T-b'}\) (red curves), which is similar to Eqn.~(\ref{theta_and_D}) but with the exponent of $T$ allowed to vary and \(\theta = K \sqrt{T-b}\) (i.e.~Eqn.~(\ref{theta_and_D})) (orange curves). The $x$-intercepts, \(b'\) and \(b\), which represent the thickness of the Kapton sheet are allowed to vary in both fits since it was challenging to retrieve the precise Kapton thickness due to the wide point spread function (PSF) of the detector. Note that the data points in the 6.2~{\textmu}m samples with thickness larger than 2.4 mm were excluded during the fitting, since they only have a small number of pixels in each bin compared to the rest of the sample, which makes those data points less reliable. From Fig.~\ref{fig:theta_thickness_fitted}, we can see that the red curves, which assume the anomalous coefficient, \(\alpha\), to be non-zero, provide a better fit to the data points than the orange curves. This suggests the potential presence of anomalous diffusion in the sample. We also observed that the exponent of $T$ is different for each sample, ranging from \(\frac{1}{1.4}\) to \(\frac{1}{1.8}\). However, the relationship between the exponent or \(\alpha\) and the sample microstructure size $S$ is difficult to test with this many free variables in the fit. We therefore focus on the orange curves which assume \(\alpha = 0\) and relate the fitted coefficient, \(K\), to the size of the microspheres as shown in Fig.~\ref{fig:coeff_against_diameter}, which is consistent with a model where the effective scattering angle, \(\theta\), is inversely proportional to \(\sqrt{S}\) as described in Eqn.~(\ref{theta_and_D}). From Fig.~\ref{fig:coeff_against_diameter} (a), we observe an unexpectedly low value for \(K\) associated with a larger uncertainty value for the 6.2~{\textmu}m sample, but also greater uncertainty in that measure, due to the presence of air gaps, as we explain in the \hyperref[s:speckle_pattern]{Speckle pattern in the dark-field images} section of the discussion. 

\subsection*{Single-exposure quantitative dark-field imaging} 
The technique described here can extract quantitative measurements from a single sample exposure, provided that the sample-to-detector propagation distance is (i) not so short as to provide insufficient signal, and (ii) not so long as to saturate the dark-field signal. In light of this tradeoff, this section examines the optimum distance for single-exposure imaging. 

\begin{figure}[hbt!]\label{theta_mid_dist}
\centering
\includegraphics[width=0.8\linewidth]{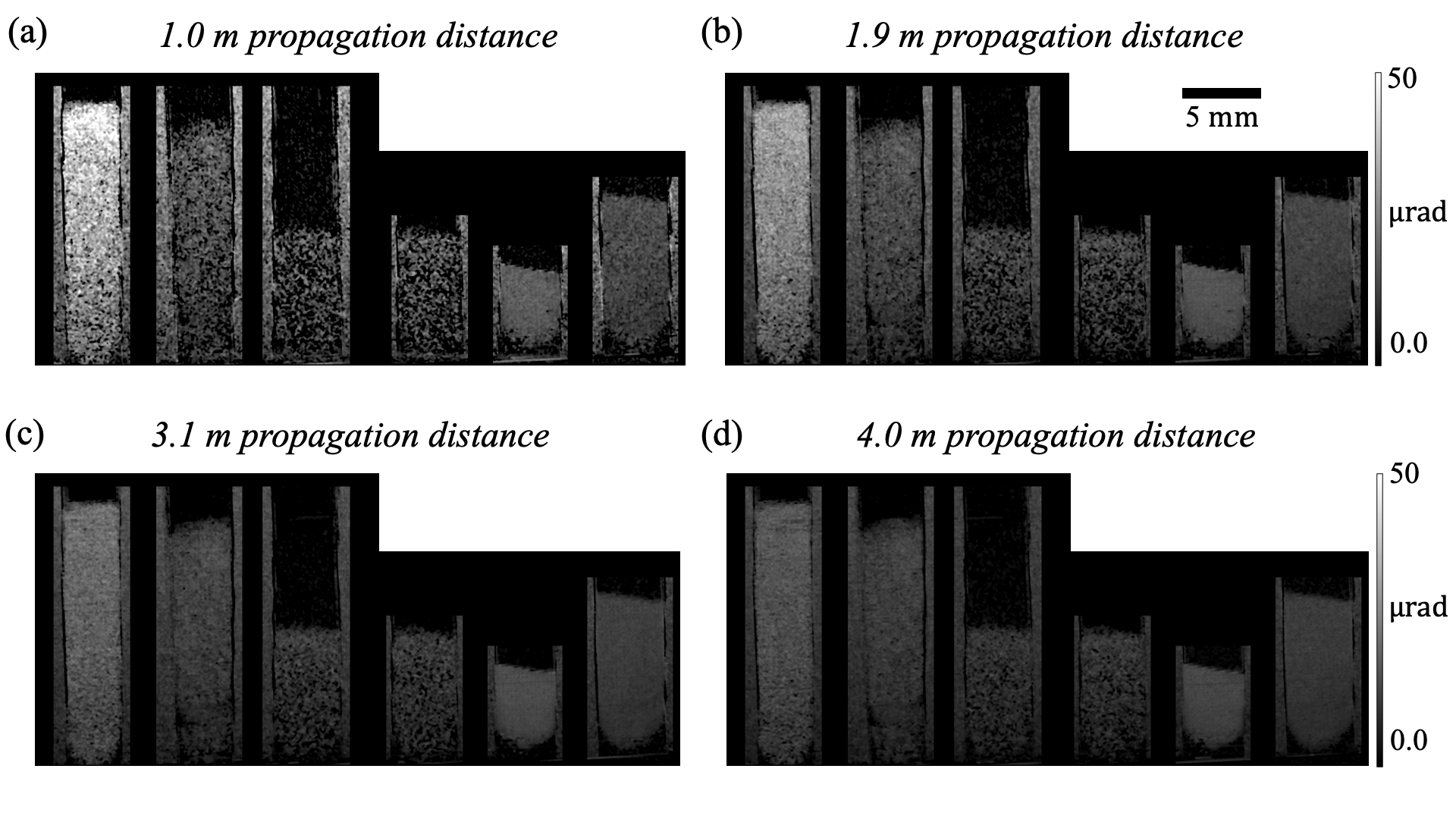}
\caption{Dark field scattering angle obtained at (\textbf{a}) 1.0 m, (\textbf{b}) 1.9 m, (\textbf{c}) 3.1 m and (\textbf{d}) 4.0 m, using a single exposure. At 1.0 m, only the scattering angle retrieved for the 1.0~{\textmu}m sample was consistent with the angle retrieved from multiple distances (Fig.~\ref{fig:combine} (m)), whereas the scattering angle retrieved for the 4.1~{\textmu}m sample in (b) and the other samples in (b) and (c) are consistent with Fig.~\ref{fig:combine} (m) and the scattering angle retrieved in (d) is smaller compared to Fig.~\ref{fig:combine} (m). At larger propagation distances ((c) \& (d)), the retrieved dark-field scattering angle image has a lower noise level compared to the images retrieved at shorter propagation distances ((a) \& (b)). Although the magnitude of the scattering angle from the 1.0~{\textmu}m sample decreases as the propagation distance increases, the measured scattering angle remains relatively consistent for the other samples between distances, despite having different noise levels. This implies our algorithm is robust in extracting quantitative dark-field signals, even when the propagation distance is not optimal.} 
\label{fig:theta_mid_dist}
\end{figure}

Figure~\ref{fig:theta_mid_dist} shows the scattering angle extracted from the single-exposure dark-field signal obtained at 4 propagation distances. The scattering angle image retrieved from a shorter propagation distance (Fig.~\ref{fig:theta_mid_dist} (a)) has significantly more noise compared to the image retrieved from a larger propagation distance (Fig.~\ref{fig:theta_mid_dist} (d)) due to the weaker dark-field sensitivity at a shorter propagation distance. The scattering angle extracted from a shorter propagation distance (Fig.~\ref{fig:theta_mid_dist} (a)) also has a greater magnitude compared to the scattering angle extracted from a larger propagation distance (Fig.~\ref{fig:theta_mid_dist} (d)). The same trends and observations are echoed in Fig.~\ref{fig:combine}. This suggests that the effective scattering angle may be overestimated at a propagation distance shorter than the optimal distance, and underestimated at a propagation distance larger than the optimal distance. 

It is worth noting that the scattering angle from the 1.0~{\textmu}m sample in Fig.~\ref{fig:theta_mid_dist} (a) is consistent with the angle retrieved from multiple distances (Fig.~\ref{fig:combine} (m)), indicating that 1.0 m is a suitable propagation distance to extract quantitative dark-field signal for this sample. Similarly, the scattering angle extracted from the other samples at 2.5 m (Fig.~\ref{fig:combine} (k)), 1.9 m and 3.1 m (Fig.~\ref{fig:theta_mid_dist} (b) and (c)) are also consistent with Fig.~\ref{fig:combine} (m), which implies the robustness of our technique towards the distance at which we perform single-exposure dark-field imaging. 

While there is some range of propagation distances that give qualitatively similar images, and a substantial range within that set that produce images that agree quantitatively, the choice of distance is important. In addition, it should be considered how the optimal single-exposure propagation distance changes for microstructures of different sizes. 

\begin{figure}[hbt!]
\centering
\includegraphics[width=0.65\linewidth]{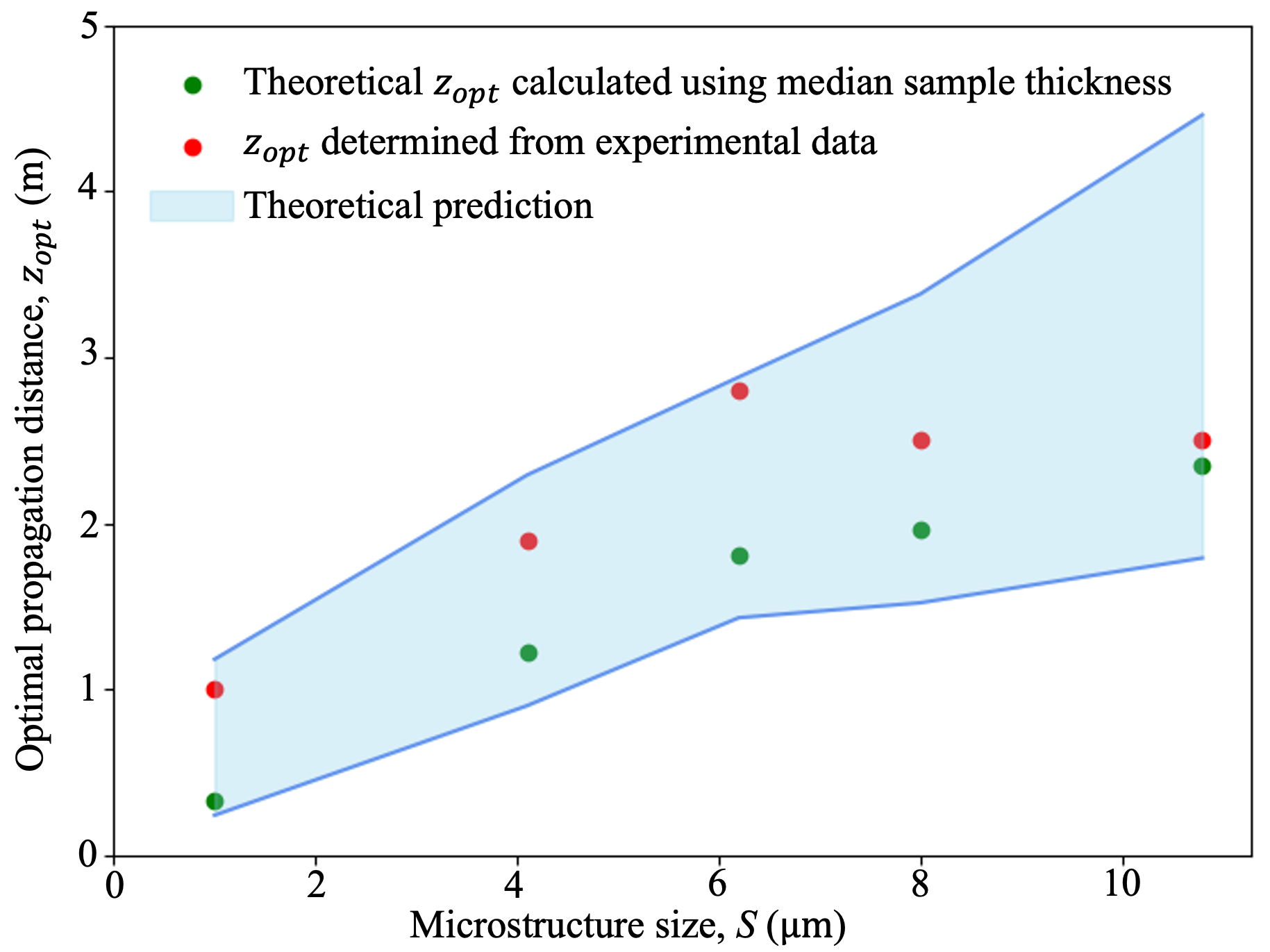}
\caption{Theoretical and measured optimal propagation distances to perform single-exposure quantitative dark-field imaging on samples with various microstructure sizes. The theoretical optimal propagation distance is calculated using Eqn.~(\ref{maths_opt_distance}) with \(T = \) median thickness, while the range of predicted viable propagation distances is obtained with \(T = \) median thickness \(\pm\) 2 standard deviations in thickness. The experimental data (see Fig.~\ref{fig:single_dist_compare} for the plot of each sample) is in agreement with the theoretical prediction.}
\label{fig:opt_dist_size}
\end{figure}

\begin{figure}[hbt!]
\centering
\includegraphics[width=0.8\linewidth]{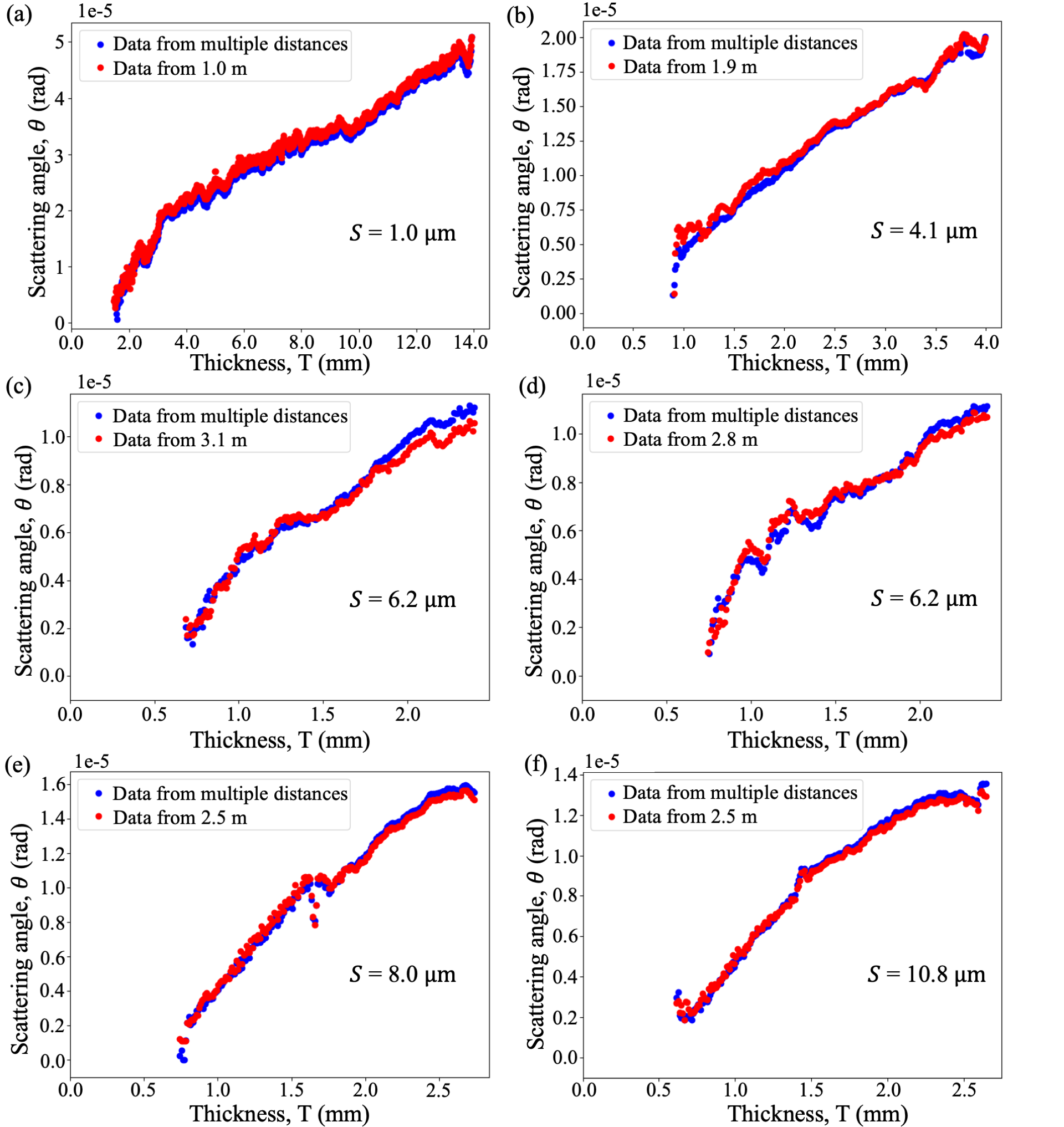}
\caption{Plots showing the dark-field scattering angle measured at the optimal propagation distance (red) using a single exposure, compared to the scattering angle extracted from multiple distances (blue) for each sample. In all panels, the recovered scattering angle is plotted against the retrieved sample thickness. Note that the optimal propagation distances chosen here are the distances at which the two curves match the best. The two curves in all samples were in agreement with each other.  Moreover, the optimal propagation distance for each sample falls within the range of theoretical prediction (see Fig.~\ref{fig:opt_dist_size}).} 
\label{fig:single_dist_compare}
\end{figure}

We can use this experimental dataset to test the optimal distance formula derived earlier in this paper. To do this, we substituted the median thickness and median \(\pm\) 2 standard deviations of the thickness of each sample into Eqn.~(\ref{maths_opt_distance}), to calculate an optimal distance and a range of suitable distances for single-exposure quantitative dark-field imaging for each sample, shown in green and blue respectively in Fig.~\ref{fig:opt_dist_size}, together with the optimal distance determined from the experimental data, shown in red. The experimental optimal distance is determined to be the distance at which the scattering angle, retrieved at such a distance using a single exposure, matches best with the scattering angle retrieved from multiple distances: see Fig.~\ref{fig:single_dist_compare}. The optimal propagation distance obtained from the experimental data lies within the range of the theoretical optimal propagation distance calculated using Eqn.~(\ref{maths_opt_distance}), showing that our result agrees with the theory. Figure~\ref{fig:opt_dist_size} shows that to extract a quantitative dark-field signal, a sample with a larger microstructure size or smaller thickness needs to be imaged at a larger propagation distance, while a sample with a smaller microstructure size or larger thickness needs to be imaged at a shorter propagation distance, as described in Eqn.~(\ref{maths_opt_distance}). However, this is not evident for the 8.0~{\textmu}m and 10.8~{\textmu}m experimental datasets, where both were found to match the multiple-distance data best at the same experimental optimal distance, 2.5 m, which is smaller than the optimal distance for the 6.2~{\textmu}m sample. This may be due to the reference grid pattern being blurred out significantly at larger propagation distances due to source-size blurring, leaving a lower visibility reference pattern to image with. This effect, not incorporated in the optimal-distance equation, renders the cross-correlation less robust to noise and background intensity variations. A lower-visibility reference pattern is more likely to result in a saturated dark-field signal at a shorter distance, raising the `visibility floor' indicated in Fig.~\ref{fig:DF_dist}.   

\section*{Discussion}
We have extracted quantitative dark-field signals from polystyrene microspheres of 5 different sizes and related the effective scattering angle to the number of microstructures. The scattering angle and the number of microstructures of all samples follows a relationship that deviates either slightly or negligibly from the theoretical square-root relationship proposed by von Nardroff\cite{von1926refraction}. As expected, it was observed that the sample with smaller microstructures produces a stronger dark-field signal and thus a larger effective scattering angle. Our theoretical model predicts that the scattering angle is inversely proportional to the microstructure size, and our data is consistent with this model. 
We have also determined a formula for the optimal propagation distance to perform quantitative single-exposure dark-field imaging. This was achieved by analytically solving for the distance at which the change in the dark-field visibility signal with respect to the change in scattering angle is maximised, and was confirmed with experimental data. 

The saturation of the dark-field signal at a larger propagation distance was observed for all samples, and depending on the sample microstructure size, the dark-field signal began saturating at a different propagation distance. Samples with smaller microstructures produced a saturated dark-field from a shorter propagation distance than the samples with larger microstructures, for equivalent thickness. 

As noted for the case of the lower-visibility reference pattern seen at large grid-to-detector distances, the visibility of the reference pattern relative to noise and background variations also affects the distance at which saturation is first observed. Therefore, within this discussion, we first examine the origin of the speckle pattern observed from the sample and how that affects the extraction of the quantitative dark-field signal. We then explore the properties of this single-exposure quantitative dark-field imaging technique, future research directions and potential applications. 

\subsection*{Speckle pattern in the dark-field images}
\label{s:speckle_pattern}
\begin{figure}[hbt!]
\centering
\includegraphics[width=0.8\linewidth]{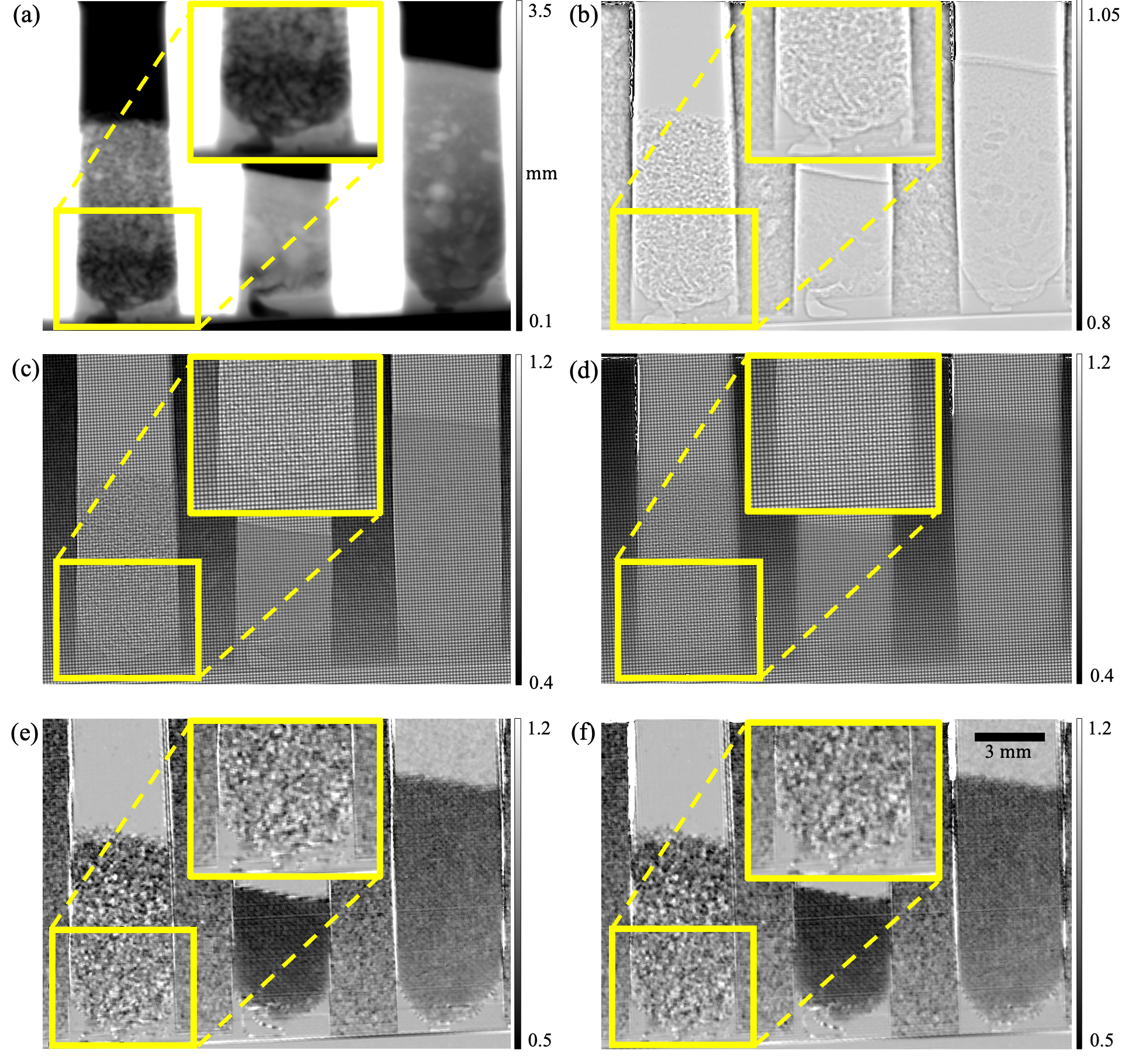}
\caption{Dark-field retrieval while removing propagation-based phase contrast edge effects. The diameter of the microspheres in each panel (from left to right) is 6.2~{\textmu}m, 8.0~{\textmu}m and 10.8~{\textmu}m. (\textbf{a}) Thickness image retrieved via the TIE-based phase retrieval algorithm \cite{paganin2002simultaneous} using the transmission signal extracted from images taken at a propagation distance of 2.8 m. (\textbf{b}) Simulated intensity of an x-ray wavefield seen after propagating a uniform-intensity x-ray wavefield by 2.8 m, via the angular spectrum approach\cite{paganin2006}, using the phase information from (a) and assuming the sample is not attenuating. (\textbf{c}) The raw sample-grid image at 2.8 m,  and (\textbf{d}) the same, after dividing out the propagation-based phase contrast edge effects (panel (b)). (\textbf{e}) \& (\textbf{f}) The dark-field signal extracted from panels (c) and (d), respectively. Some artefacts can be observed near the edges of the rubber wedges in (b). The dark-field signals in (e) and (f) are similar, suggesting that the speckle pattern observed in the dark-field-signal image does not originate from the propagation-based phase contrast fringes, but truly from local variations in dark-field.  This result also suggests that our algorithm can still extract the dark-field signal accurately, even in the presence of propagation-based phase contrast edge effects.} 
\label{fig:pbi_cor_figure}
\end{figure}

From Fig.~\ref{fig:combine}, we observed that the 6.2~{\textmu}m sample produced stronger local variations in the change in grid visibility, in the mean (transmission) intensity, and hence the dark-field signal and scattering angle images, compared to the rest of the sample. The speckle pattern observed in the dark-field image of the 6.2~{\textmu}m sample may be attributed in part to the propagation-based phase contrast edge effects, contrast which is evident from the transmission images (see Fig.~\ref{fig:combine} (d), (e) \& (f)), increasing with propagation distance. If this were the case, then by removing the propagation-based phase effect, we should retrieve a dark-field signal image with no speckle pattern. This correction can be achieved by dividing the raw data by the simulated intensity of the x-ray wavefield propagated through free space after exiting from the sample (as demonstrated in Groenendijk \textit{et al}.~\cite{groenendijk2020material}), before performing our method of analysis. Figure~\ref{fig:pbi_cor_figure} (b) shows the resulting intensity seen after propagating an x-ray wavefield with uniform intensity through the sample (with thickness shown in Fig.~\ref{fig:pbi_cor_figure} (a)) and then an extra distance of 2.8 m, with the free-space propagation performed via the angular-spectrum approach \cite{paganin2006} with an assumption that the sample is not attenuating, using \(\delta = 2.0463 \times 10^{-7}\) (polystyrene at 34keV) and \(\beta = 0\), where the complex refractive index is \(n = 1 - \delta + i\beta\). Figure~\ref{fig:pbi_cor_figure} (d) shows the result obtained after dividing the raw sample-grid image (Fig.~\ref{fig:pbi_cor_figure} (c)) by the propagated grid-free intensity determined by simulation (Fig.~\ref{fig:pbi_cor_figure} (b)). The majority of the fringes formed within the sample itself and near the edges of the glue down the bottom and near the rubber wedges can no longer be seen in Fig.~\ref{fig:pbi_cor_figure} (d), indicating that the propagation-based phase effects have been removed. 
 
The dark-field signals extracted from the images taken at a propagation distance of 2.8 m before and after this correction are shown in Fig.~\ref{fig:pbi_cor_figure} (e) and (f) respectively. We observed that Fig.~\ref{fig:pbi_cor_figure} (f) agrees with Fig.~\ref{fig:pbi_cor_figure} (e) except that (f) appears to be very slightly smoother. This not only suggests that the speckle pattern formed in the 6.2~{\textmu}m sample tube does not originate from the propagation-based phase effects, but it also indicates that our algorithm can still work effectively even in the presence of propagation-based fringes. (Images of other samples after correction, and related relevant results, can be found in Fig.~S1 from the Supplementary Information.)

The presence of clusters and air gaps in the 6.2~{\textmu}m tube, as described earlier in the paper, can be further justified by examining Fig.~\ref{fig:pbi_cor_figure} (c) \& (d).  Here, we see that some regions in the grid intensity pattern are more blurred out, compared to the rest of the pattern.  As expected, this corresponds to the regions with greater magnitude in the thickness image (Fig.~\ref{fig:pbi_cor_figure} (a)) and also to the regions with stronger dark-field signal in Fig.~\ref{fig:pbi_cor_figure} (e) \& (f). It is also worth noting that the clusters formed in the 6.2~{\textmu}m sample also act as additional microstructure in the sample, with a length scale different from the individual microspheres. The discontinuity observed from the dark-field signal produced by the 6.2~{\textmu}m sample across different propagation distances (see Fig.~S2 in the Supplementary Information) suggests that the blurring from the extra sample microstructure may be of the same length scale as the grid period, thus causing the grid intensity pattern to distort further, until the cross-correlation analysis could not work properly and potentially also leading to the saturation of the dark-field signal. 

A small number of pixels in the 6.2~{\textmu}m sample tube have shown an increase in visibility, as demonstrated by the dark-field signals detected from these pixels being greater than 1. This may be due to the focusing of x-rays by the air gaps. It may also be due to clumps within the sample that are barely resolved -- i.e., not sufficiently smaller relative to the pixel size to produce a dark-field signal, but also not sufficiently well-resolved to produce a visible phase shift in the x-ray wavefield. Another contributing factor may be grid-period-sized phase-induced distortions coming from the barely-resolved sample clumps mentioned above, which may locally change the visibility of the reference intensity pattern.   

\subsection*{Properties of the technique} 
\label{s:properties}
One main advantage of this technique is that it only requires a single sample exposure to extract the attenuation, phase-shift and dark-field signal. The short data acquisition time can minimise the radiation dose delivered to the sample and the potential motion blurring in the image, which makes this technique more feasible for dynamic imaging or time-sequence imaging, compared to other PCXI techniques that need multiple sample exposures to extract dark-field images. This technique also has a relatively simple setup, compared to other dark-field imaging techniques. It does not need any additional optical elements other than an attenuating or phase grid. Furthermore, no calibration or alignment is required for this technique, prior to the data acquisition process. The sensitivity of the imaging system to the dark-field signal can also be tuned, by changing the pixel size and the propagation distance. We can increase the sensitivity by using a smaller pixel size or a larger propagation distance. 

The single-grid imaging technique can also be performed in a laboratory-based setup using polychromatic x-ray sources with non-negligible source size \cite{wen2010,macindoe2016}, since it only requires a certain degree of spatial coherence of the source \cite{macindoe2016}, which could be improved by placing a grid immediately downstream of the source. Although the single-grid imaging technique only has a weak requirement on the temporal coherence of the source, the polychromaticity of the source could be an issue for quantitative dark-field imaging, due to the beam hardening effect, which can contribute to a `pseudo' dark-field signal \cite{chabior2011beam}. 

Our technique is robust with respect to propagation distance, with the effective scattering angle extracted from a single exposure consistent within a range of propagation distances, centred on the optimal distance. The optimal distance to perform single-exposure dark-field imaging for a sample with estimated thickness and microstructure size can be determined using Eqn.~(\ref{maths_opt_distance}), where the value of \(m\) can be obtained from a calibration curve of samples made up of the same material, with known thickness and size. A simple general rule---in choosing a suitable sample-to-detector distance experimentally---is to image the sample at a propagation distance that is large enough such that the blurring from the sample can be observed clearly in the intensity pattern formed on the detector, but not so large that the intensity pattern is completely blurred out. 

Another advantage of this technique is that the spatial-mapping/explicit-tracking method \cite{morgan2011_grid} we applied here can provide a higher spatial resolution than a Fourier-analysis approach \cite{takeda1982}, since we are comparing the windows pixel-by-pixel. Such Fourier analysis can fail when the grid frequency overlaps with the sample feature frequency (as shown in Fig. 4 from Morgan {\em et al.}~\cite{morgan2013}). Although the results from Morgan {\em et al.}~\cite{morgan2013} focused on phase imaging, we believe the same applies for dark-field imaging. This spatial-mapping approach also allows the algorithm to be successfully applied on images taken using speckle-based imaging, as shown in Section 6.2 of How \& Morgan~\cite{how2022quantifying}, where the grid is replaced with a piece of sandpaper \cite{berujon2012_sb,morgan2012}.

However, the single-grid imaging technique has some limitations. One major limitation is that this technique is primarily suitable for millimetre-to-centimetre-sized samples. This is due to the fact that it requires a relatively small pixel size to capture the changes to the grid pattern directly. To obtain high-quality dark-field images, the pixel size usually has to be smaller than the blurring width introduced by the sample. Although technically this technique can be used to image a large sample by using a detector with a larger area but a sufficiently small pixel size, this can introduce a higher radiation dose on the sample and may not be feasible due to the cost and/or lack of availability of such a detector.  

\subsection*{Potential applications and future directions} 
As explained in the previous section, this technique can be useful for imaging dynamic or irreversible processes, such as \textit{in vivo} biological response to a treatment \cite{morgan2014vivo}, chemical reactions \cite{rieger2000study,albiter2003structural} and the production of metal foams \cite{garcia2019using}. This technique can also be extended into three dimensions to acquire fast tomography, which can be useful for biomedical applications, such as lung imaging \cite{yaroshenko2013pulmonary,kitchen2020emphysema}, breast tissue imaging for early detection of cancer \cite{michel2013dark}, and kidney stone imaging for classification \cite{scherer2015non}. In particular, the dark-field signal can provide information about the size of alveoli \cite{yaroshenko2013pulmonary,kitchen2020emphysema}, and thus this technique has the potential to be used as a diagnostic tool for lung diseases such as emphysema \cite{hellbach2015vivo}, fibrosis \cite{hellbach2017x}, chronic obstructive pulmonary disease (COPD) \cite{willer2021x}, and lung cancer \cite{scherer2017x}, which can produce changes in the size and structure of the alveoli. This technique can also be used to study chemical reactions that involve the forming or decomposition of substances in an aqueous solution, which act as unresolved microstructures that provide the dark-field signals \cite{rieger2000study,albiter2003structural}. Moreover, by quantifying the dark-field signal and relating the signal to the sample microstructure size and material, this technique can also be used to identify powder-like goods, which can be useful for airport security to detect explosive substances or powdered drugs \cite{miller2013phase}.    

In this manuscript, we have extracted the effective scattering angle from the dark-field signal using multiple distances, and related this angle to the sample microstructure size. However, this does not fully explain the relationship between the dark-field signal and the scattering from sample microstructures. It would be interesting to further understand how the dark-field signal relates quantitatively to the small-angle or ultra-small-angle x-ray scattering (SAXS or USAXS).  This model can be further improved by investigating how the dark-field signal is related to the statistical moments that describe the shape of the SAXS or USAXS distribution \cite{modregger2012, modregger2017}. It would also be interesting to determine the fraction of the propagating x-ray wavefield that contributes to the dark-field signal \cite{paganin2019x}.

In applying the method to a range of sample sizes, it may be important to separate the edge-related phase effects from the dark-field effects. In some regimes, if a phase fringe has a length scale comparable to the reference grid period, the fringe could be misinterpreted as a dark-field signal, although we have not seen that here (Fig.~\ref{fig:pbi_cor_figure}). Future work could investigate how the strength of the dark-field signal is determined by the sample material and microstructure shape, and how this relates to the optimal propagation distance to perform quantitative single-exposure dark-field imaging.     

Another direction is to investigate quantities associated with directional dark-field signal\cite{jensen2010a,jensen2010b}, which comes from elongated microstructures that are oriented in a certain direction. By modelling the blurring function as a two-dimensional Gaussian distribution\cite{jensen2010b}, Croughan \textit{et al}. \cite{croughan2022directional} have successfully applied this technique to extract the directional dark-field signal, including both the angle at which the microstructures are oriented and the eccentricity of the blurring from the microstructures. This quantitative directional dark-field imaging technique can be used to determine the diameter of fibres, which have a length many times their width, and hence a scattering angle much larger in one direction. 

\section*{Conclusion}
This manuscript has derived and tested guidelines for performing single-exposure dark-field x-ray imaging to quantify the size of dark-field-generating microstructure. To provide experimental evidence, we have extracted the diffusive dark-field signal from polystyrene microspheres of diameter 1.0~{\textmu}m, 4.1~{\textmu}m, 6.2~{\textmu}m, 8.0~{\textmu}m and 10.8~{\textmu}m, using  multiple propagation distances, employing the single-grid algorithm developed by How \& Morgan \cite{how2022quantifying}. We observed that a sample with smaller microstructures produces a stronger dark-field signal, compared to a sample with larger microstructures of the same projected thickness. The retrieved dark-field scattering angle was consistent with our theoretical model stating the angle is inversely proportional to the square root of sample microstructure size. 

We determined an expression for the optimal sample-to-detector distance range for single-exposure dark-field imaging, by analytically solving for the distance at which the change in dark-field visibility signal with respect to the change in scattering angle is maximised. This also avoids both the insufficient signal seen at short sample-to-detector distances and the dark-field saturation seen at large distances. According to the analytical solution, the optimal distance for single-exposure dark-field imaging depends on the grid period, sample microstructure size and sample thickness. We have verified the theoretical model by comparing the effective scattering angle extracted from a single exposure at a single distance, to the angle extracted from multiple distances.  Here, the experimental optimal distance falls within the theoretical optimal distance range, for all samples. 

Single-grid dark-field imaging can characterise sample microstructure, once calibrated with known microstructures of comparable material and size to those we wish to investigate. Imaging can then be performed with a single sample exposure, allowing time-sequence or low-dose imaging. 

\section*{Data availability} 
Data underlying the results presented in this paper is not publicly available at this time but may be obtained from the authors upon reasonable request. 

\bibliography{reference_list}

\section*{Acknowledgements}
We acknowledge ARC Future Fellowship funding (FT180100374). The images in this paper were captured at the Imaging and Medical Beamline at the Australian Synchrotron, part of ANSTO, under application 16818a. We also thank Michelle Croughan for providing constructive feedback on the manuscript.  

\section*{Author contributions} 
Y.Y.H and K.S.M. designed the technique. Y.Y.H and K.S.M. performed the experiment and wrote the analysis code. The paper was mainly written by Y.Y.H., with input from K.S.M. and D.M.P., who also provided guidance and supervision of the project. 

\section*{Competing interests}
The author(s) declare no competing interests.

\end{document}


\flushbottom
\maketitle

\thispagestyle{empty}

\noindent Two supplementary figures are given here.  These are referred to in the section of the main paper that is titled `Speckle pattern in the dark-field images'. 

\begin{figure}[hbt!]
\centering
\includegraphics[width=0.7\linewidth]{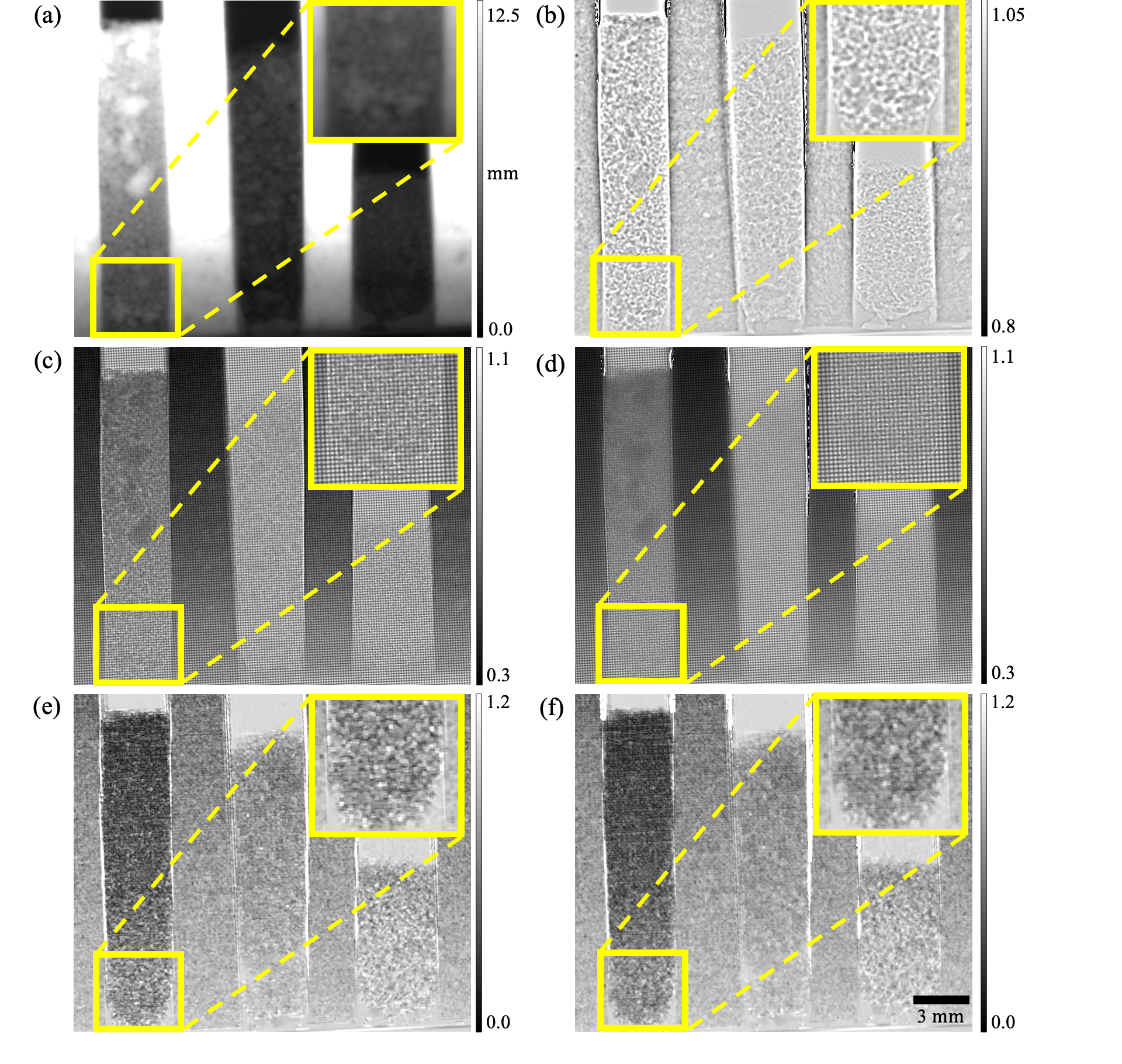}
\caption{Dark-field retrieval, using an analysis which removes edge effects that are due to propagation-based phase contrast. The diameter of the microspheres in each panel (from left to right) is 1.0~{\textmu}m, 4.1~{\textmu}m and 6.2~{\textmu}m, respectively. (\textbf{a}) Thickness image recovered via a phase-retrieval algorithm\cite{paganin2002simultaneous} that is based on the transport-of-intensity equation\cite{Teague1983}, using the transmission signal extracted from images taken at a propagation distance of 2.8 m. (\textbf{b}) Simulated intensity of an x-ray wavefield with uniform intensity, after being numerically propagated 2.8 m via the angular spectrum approach\cite{paganin2006}, using the phase information from (a) and assuming the sample is not attenuating. (\textbf{c}) The raw sample-plus-grid image at 2.8 m,  and (\textbf{d}) the same, after dividing out the propagation-based phase contrast edge effects (panel (b)). (\textbf{e}) \& (\textbf{f}) The dark-field signal extracted from panels (c) and (d), respectively. Some artefacts can be observed near the edges of the rubber wedges in (b).} 
\end{figure}

\begin{figure}[hbt!]
\centering
\includegraphics[width=0.45\linewidth]{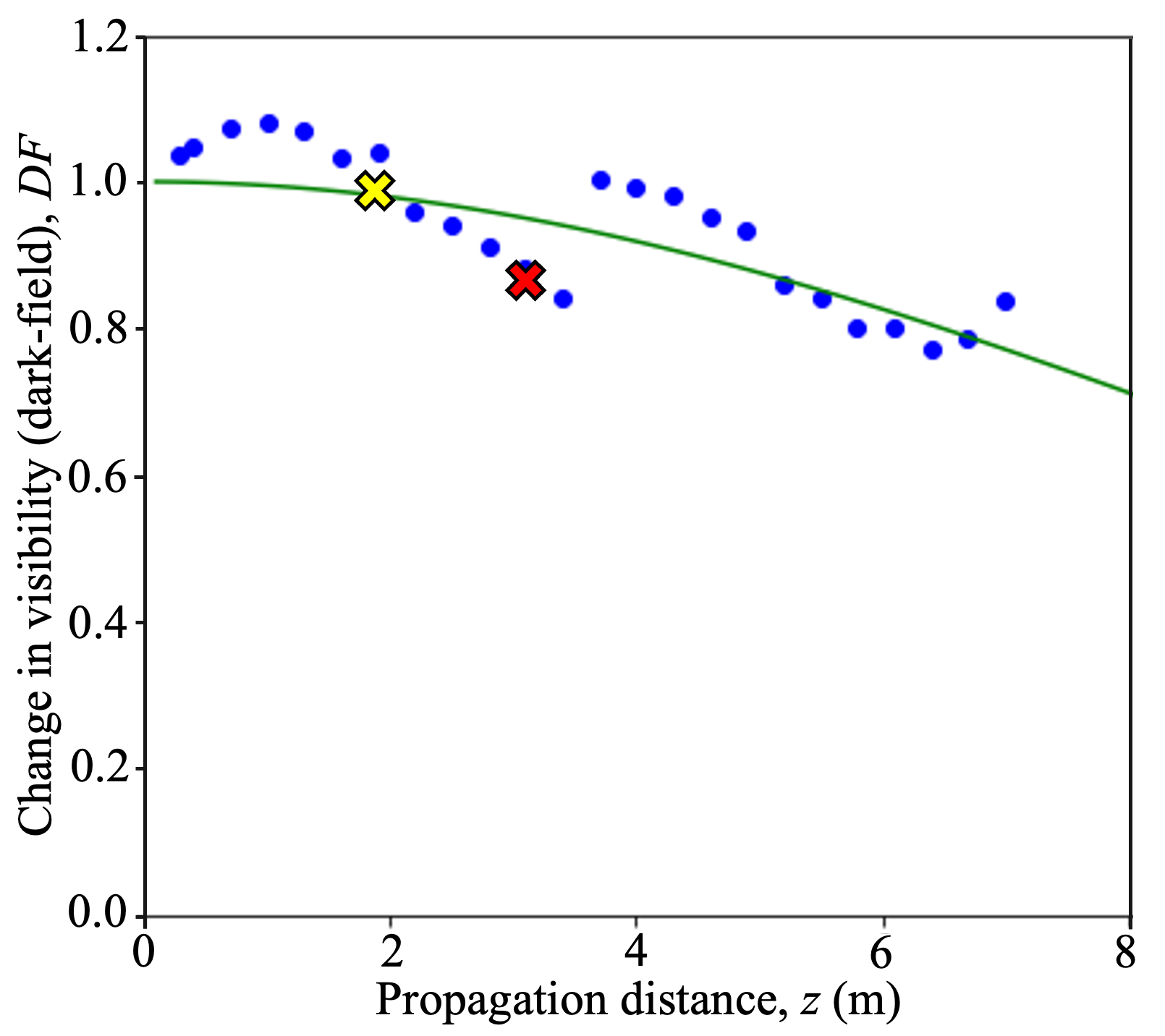}
\caption{A typical example of the dark-field signal measured as a loss in visibility (at pixel (960, 490), from the 6.2~{\textmu}m sample) using 24 propagation distances. The red and yellow crosses label the dark-field saturation and maximum-gradient point, respectively, as explained in Fig.~1 of the main text of the paper. One or multiple discontinuities are observed from the dark-field signal measured from this sample across different propagation distances. This may be due to the extra distortion of the grid intensity pattern, contributed by blurring from the extra sample microstructure of the same length scale as the grid period. The scattering angle is extracted from the dark-field signal, measured at the shorter propagation distances to minimise the contributions from the extra sample microstructure.}
\end{figure}

\bibliography{reference_supp}